\documentclass[letterpaper]{article} 
\usepackage{aaai19}  
\usepackage{times}  
\usepackage{helvet} 
\usepackage{courier}  
\usepackage[hyphens]{url}  
\usepackage{graphicx} 
\usepackage{graphicx}  
\usepackage{latexsym}

\usepackage[utf8]{inputenc}
\usepackage{amsmath,amsfonts,amssymb}
\usepackage{mathtools}
\usepackage{stmaryrd}
\usepackage{listings}
\usepackage{microtype}

\usepackage{enumitem}
\usepackage{xcolor}
\usepackage{xspace}

\binoppenalty=\maxdimen
\relpenalty=\maxdimen

\newcommand{\citet}[1]{\citeauthor{#1} \shortcite{#1}}
\newcommand{\citep}{\cite}
\newcommand{\citealp}[1]{\citeauthor{#1} \citeyear{#1}}

\nocopyright 

\begin{document}

\title{A Uniform Treatment of Aggregates and Constraints in Hybrid ASP}

\author{%
  Pedro Cabalar$^1$\and
  Jorge Fandinno$^2$\and
  Torsten Schaub$^2$\and
  Philipp Wanko$^2$\\
  $^1$ University of Corunna, Spain\\
  $^2$ University of Potsdam, Germany}

\newtheorem{definition}{Definition}
\newtheorem{proposition}{Proposition}
\newtheorem{corollary}{Corollary}
\newtheorem{observation}{Observation}
\newtheorem{lemma}{Lemma}
\newtheorem{example}{Example}
\newtheorem{theorem}{Theorem}

\newenvironment{proof}[1]{\paragraph{Proof of {#1}}}{\hfill$\square$\\}

\newcommand{\den}[1]{\llbracket \, #1 \, \rrbracket}
\newcommand{\ctermm}[3]{\ensuremath{{#1|#2}{:\,}#3}}
\newcommand{\cterm}[3]{\ensuremath{(\ctermm{#1}{#2}{#3})}} 
\newcommand{\close}[1]{\ensuremath{#1\raisebox{1pt}{$\uparrow$}}}

\newcommand{\htag}[2]{\ensuremath{\mathit{#1}\agg{#2}}}
\newcommand{\htagg}[4]{\ensuremath{\mathit{#1}\agg{#2}\mathrel{#3}#4}}
\newcommand{\htaggg}[4]{\ensuremath{\mathit{#1} {#2 }\mathrel{#3}#4}}
\newcommand{\vals}[2]{\ensuremath{\mathcal{V}_{#1,#2}}}
\newcommand{\eval}[2]{\ensuremath{\mathit{eval}}_{\langle #1,#2\rangle}}

\newcommand{\evalgz}[2]{\ensuremath{{\gz}_{\langle #1,#2\rangle}}}
\newcommand{\evalf}[2]{\ensuremath{{\f}_{\langle #1,#2\rangle}}}
\newcommand{\evals}[3]{\ensuremath{{#1}_{\langle #2,#3\rangle}}}
\newcommand{\evalcl}[1]{\ensuremath{\mathit{eval}_{#1}}}
\newcommand{\evall}[1]{\ensuremath{\mathit{eval}}_{#1}}
\newcommand{\evallgz}[1]{\ensuremath{\mathit{eval}^{\mathtt{gz}}_{#1}}}

\newcommand{\modelscl}{\ensuremath{\models_{cl}}}
\newcommand{\eqdef}{%
  \mathrel{\vbox{\offinterlineskip\ialign{%
    \hfil##\hfil\cr%
    $\scriptscriptstyle\mathrm{def}$\cr%
    \noalign{\kern1pt}%
    $=$\cr%
    \noalign{\kern-0.1pt}%
}}}}

\newcommand{\sysfont}{\textit}
\newcommand{\clingo}{\sysfont{clingo}}
\newcommand{\clingoDL}{\clingo{\small\textnormal{[}\textsc{DL}\textnormal{]}}}
\newcommand{\dlprogram}{\textit{DL}-program}
\newcommand{\code}[1]{\texttt{#1}}
\newcommand{\ground}{variable-free}
\newcommand{\modelsgz}{\models_{\gz}}
\newcommand{\modelsf}{\models_{\f}}
\newcommand{\modelss}{\models_{\sem}}
\newcommand{\modelsp}[1]{\models_{#1}}
\newcommand{\Atoms}{\mathit{Atoms}}
\newcommand{\Head}{\mathit{H}}
\newcommand{\HeadA}{\mathit{\Head_A}}
\newcommand{\HeadB}{\mathit{\Head_B}}
\newcommand{\Headp}{\mathit{\Head^+}}
\newcommand{\Headn}{\mathit{\Head^-}}
\newcommand{\Body}{\mathit{B}}
\newcommand{\Bodyp}{\mathit{\Body^+}}
\newcommand{\Bodyn}{\mathit{\Body^-}}
\newcommand{\var}{\mathit{var}}
\newcommand{\Cond}{\mathit{Cond}}
\newcommand{\tuple}[1]{\langle #1 \rangle}
\newcommand{\restr}[2]{{#1|}_{\hspace{-1pt}#2}}
\newcommand{\df}[1]{\mathit{def}(#1)}
\newcommand{\HTC}{\ensuremath{\mathit{HT}_{\!C}}} 
\newcommand{\HT}{\ensuremath{\mathit{HT}}\xspace}
\newcommand{\X}{\ensuremath{\mathcal{X}}}
\newcommand{\D}{\ensuremath{\mathcal{D}}}
\newcommand{\Du}{\ensuremath{\mathcal{D}_{\mathbf{u}}}}
\newcommand{\T}{\ensuremath{\mathcal{T}}}
\newcommand{\C}{\ensuremath{\mathcal{C}}}
\newcommand{\ET}{\ensuremath{\mathcal{T}^e}}
\newcommand{\BT}{\ensuremath{\mathcal{T}^b}}
\newcommand{\BC}{\ensuremath{\mathcal{C}^b}}
\newcommand{\Cgr}{\ensuremath{\mathcal{C}^*}}
\newcommand{\F}{\ensuremath{\mathcal{F}}}
\newcommand{\I}{\ensuremath{\mathcal{I}}}
\newcommand{\V}{\ensuremath{\mathcal{V}}}
\newcommand{\A}{\ensuremath{\mathcal{A}}}
\newcommand{\Piref}[1]{\Pi_{\ref{#1}}}
\newcommand{\vars}[1]{\ensuremath{\mathit{vars}(#1)}}
\newcommand{\varsp}[1]{\ensuremath{\mathit{vars}^{a}(#1)}}
\newcommand{\atoms}[1]{\ensuremath{\mathit{atoms}(#1)}}
\newcommand{\At}[1]{\ensuremath{\mathit{At}(#1)}}
\newcommand{\val}{\ensuremath{v}} 
\newcommand{\gz}{\ensuremath{\mathit{vc}}} 
\newcommand{\f}{\ensuremath{\mathit{df}}} 
\newcommand{\ff}{\ensuremath{\mathit{F}}} 
\newcommand{\undefined}{\ensuremath{\mathbf{u}}} 
\newcommand{\sem}{\ensuremath{\kappa}} 
\newcommand{\ass}[3]{#1 := #2 \, .. \, #3}
\newcommand{\LC}{\ensuremath{\mathit{LC}}}
\newcommand{\DF}{\ensuremath{\boldsymbol{D\hspace{-1.2pt}F}}} 
\newcommand{\agg}[1]{\ensuremath{\dot{\{}#1\dot{\}}}} 
\newcommand{\Z}{\ensuremath{\mathbb{Z}}}
\newcommand{\LX}{\ensuremath{\mathbb{X}}}
\newcommand{\HU}{\ensuremath{\mathbb{U}}}
\newcommand{\Gra}{\ensuremath{\mathit{G}^a}}
\newcommand{\Gr}{\ensuremath{\mathit{G}}}
\newcommand{\Def}{\delta}
\newcommand{\grsep}{\,\big|\hspace{-3pt}\big|\hspace{-3pt}\big|\,}
\newcommand{\isint}[1]{\ensuremath{\mathit{int}(#1)}}

\let\olditem\item
\let\oldenumerate\enumerate
\let\oldendenumerate\endenumerate
\let\olditemize\itemize
\let\oldenditemize\enditemize
\newcommand{\deitemize}{%
\def\itemize{\def\item{}}
\def\enditemize{\let\item\olditem}
\def\enumerate{\let\item\olditem\oldenumerate}
\def\endenumerate{\oldendenumerate\def\item{}}
}

\newcommand{\reitemize}{%
\def\itemize{\olditemize}
\def\enditemize{\oldenditemize}
}


\newenvironment{Itemize}{\begin{itemize}[leftmargin=0pt]}{\end{itemize}}


\renewenvironment{Itemize}{\let\item\relax}{\let\item\olditemize}
\renewenvironment{itemize}{\let\item\olditem\olditemize}{\oldenditemize\let\item\relax}
\renewenvironment{enumerate}{\let\item\olditem\oldenumerate}{\oldendenumerate\let\item\relax}


\def\sumf{\mathit{sum}}
\newcommand\op{\mathit{op}}
\newcommand\EX{\mathcal{EX}}

\newcommand{\appendblank}{\ensuremath{\circ}}
\newcommand\theory[2]{\tuple{#1,#2}}
\newcommand{\aggs}{\mathit{agg}}
\newcommand{\levels}{\ell}
\newcommand{\level}[1]{\levels(#1)}
\newcommand{\gzsemantics}{\mbox{\gz-semantics}\xspace}
\newcommand{\fsemantics}{\mbox{\f-semantics}\xspace}
\newcommand{\sequilibrium}{\mbox{\sem-equilibrium}\xspace}
\newcommand{\gzequilibrium}{\mbox{\gz-equilibrium}\xspace}
\newcommand{\fequilibrium}{\mbox{\f-equilibrium}\xspace}
\newcommand{\sstable}{\mbox{\sem-stable}\xspace}
\newcommand{\gzstable}{\mbox{\gz-stable}\xspace}
\newcommand{\fstable}{\mbox{\f-stable}\xspace}
\def\kmapping{selection function}
\def\kmappings{{\kmapping}s}

\maketitle

\pagestyle{plain}

\begin{abstract}
  Characterizing hybrid ASP solving in a generic way is difficult since one needs to abstract from specific theories.
  Inspired by lazy SMT solving, this is usually addressed by treating theory atoms as opaque.
  Unlike this, we propose a slightly more transparent approach that includes an abstract notion of a term.
  Rather than imposing a syntax on terms, we keep them abstract by stipulating only some basic properties.
  With this, we further develop a semantic framework for hybrid ASP solving and provide aggregate functions for theory
  variables that adhere to different semantic principles, show that they generalize existing aggregate semantics in ASP
  and how we can rely on off-the-shelf hybrid solvers for implementation.
\end{abstract}


\section{Introduction}\label{sec:introduction}

Many real-world applications have a heterogeneous nature
that can only be captured by different types of constraints.
This is commonly addressed by hybrid solving technology,
most successfully in the area of Satisfiability modulo Theories (SMT;~\citealp{niolti06a}).
Meanwhile,
neighboring areas like Answer Set Programming (ASP;~\citealp{lifschitz08b}) follow suit.
In doing so, they usually adopt the lazy approach to SMT that abstracts from specific constraints
by interpreting them as opaque atoms.
This integration is however often done in system-oriented ways that leave semantic aspects behind.

We first addressed this issue in~\cite{cakaossc16a} by providing a uniform semantic framework that allows us
to capture the integration of ASP with foreign theories.
This blends the non-monotonic aspects of ASP with other formalisms in a homogeneous
representational framework.
Moreover, it retains the representational aspects of ASP such as expressing defaults and an easy
formulation of reachability, and transfers them to the integrated theory.
In~\cite{cafascwa2020a}, we extended this to conditional aggregates,
which already incurred a fraction of the aforementioned opaqueness principle.
To illustrate this, consider the following hybrid ASP rule, taken \mbox{from~\cite{cafascwa2020a}}%
\footnote{We put dots on top of braces, viz.~``$\agg{ \dotsc }$'', to indicate \emph{multisets}.}
\begin{gather*}\label{eq:tax.sum}
  \mathit{total}(R) := \mathit{sum}\agg{ \, \mathit{tax}(P) : \mathit{lives}(P,R) \,  } \ \leftarrow \ \mathit{region}(R)
  \quad
\end{gather*}
This rule gathers the total tax revenue of each region $R$ by summing up the tax liabilities of the region's residents, $P$.

The need for subatomic structures emerges from the observation
that the meaning of this rule should remain unchanged,
in case the computation of the revenue is expressed using, for instance, a linear expression instead of the $\mathit{sum}$ aggregate.
However,
this slight syntactic difference leads to a distinct constraint atom,
whose semantics can be radically different.
Only by inspecting the subatomic structure of both atoms, we can guarantee the expected behavior.

In this paper,
we build an account of such abstract subatomic structures, namely \emph{constraint terms},
and leverage them to provide a uniform treatment of linear constraints, conditional expressions, aggregates
and similar future hybrid constructs.
Furthermore,
we investigate two different principles for conditional expressions:
the \emph{vicious circle principle}~(\gz) and a new one we call \emph{definedness} (\f).
While \gz\ has been investigated in traditional ASP in~\cite{gelzha19a},
this new principle ensures that the value of any conditional expression is always defined.
This is different from \gz\ according to which conditional expressions may be undefined due to cyclic dependencies~\cite{cafafape18a}.
Interestingly,
when combined with aggregates,
the \f\ principle leads to a generalization of another semantics known from ASP~\cite{ferraris11a},
which provides the semantic underpinnings of aggregates used in the ASP system \clingo~\cite{gekakasc17a}.
Hence, for characterizing hybrid variants of \clingo, this framework is a prime candidate.
Moreover,
we are able to show how, under certain circumstances,
arithmetic aggregates (under both principles) can be mapped into
conditional linear constraints under \gz.
Combined with our previous results~\cite{cakaossc16a,cafascwa2020a},
this allows us to use off-the-shelf constraint ASP (CASP;~\citealp{lierler14a}) solvers to implement such hybrid extensions.


\section{Here-and-There with Conditional Constraints}\label{sec:approach}

The syntax of the logic \HTC\ is based on a set of (constraint) variables~$\X$
and constants or domain values
from some non-empty set~$\D$.
For convenience, we also distinguish a special symbol $\undefined \notin \X \cup \D$ that stands for \emph{undefined}.

We introduce next what we call \emph{basic} constraint terms, atoms and formulas and then extend these three concepts to incorporate conditional expressions.
We define the set of \emph{elementary terms} $\ET \eqdef {\X \cup \D \cup \{ \undefined \}}$, that is, variables, domain values and the symbol $\undefined$.
Each theory will be defined over a given set of \emph{basic (constraint) terms}, denoted as $\BT$, that will include, at least, all elementary terms, i.e., $\ET \subseteq \BT$.
The syntax of a basic term is left open, but can be any expression of infinite length.
A \emph{basic (constraint) atom} is an expression containing a (possibly infinite\footnote{An atom may have an infinite number of terms of infinite length. This cannot be represented as a string, but is still a expression similar, for instance, to some formula in infinitary logics.}) number of basic terms.
For each theory, we assume a particular set of basic constraint atoms, denoted as $\BC$.
We do not impose any limitation on their syntax,
though, in most cases, that syntax is defined by some grammar or regular pattern.
For instance, \emph{difference constraint atoms} are expressions of the form ``$x - y \leq d$'',
containing the elementary terms $x, y, d$ where $x,y \in \mathcal{X}$ are variables and $d \in \mathcal{D}$ a domain value.
Note that we are free to define the subexpression ``$x-y$'' as a basic term or not, at our convenience.
This does not affects the definition of ``$x - y \leq d$'' as a basic atom.
The importance of distinguishing terms is thus not syntactic,
but meta-logical:
Distinguishing terms allows us to guarantee some properties that may not be satisfied on unstructured atoms.

A \emph{basic formula} $\varphi$ over $\BC$ is defined as
\begin{align*}
\varphi::= \bot \mid c\mid \varphi \land \varphi \mid  \varphi \lor \varphi \mid  \varphi \rightarrow \varphi \quad\text{ where }c\in \BC
\end{align*}
We define $\top$ as ${\bot \rightarrow \bot}$ and ${\neg\varphi}$ as ${\varphi \rightarrow \bot}$ for every formula~$\varphi$.
We sometimes write ${\varphi \leftarrow \psi}$ instead of ${\psi \to \varphi}$
to follow logic programming conventions.

We now extend these notions to incorporate conditional constructs.
A \emph{conditional term} is an expression of the form~
\[
\cterm{s}{s'}{\varphi}
\]
where~$s$ and~$s'$ are basic terms and~$\varphi$ is a basic formula.
The intuitive reading of a conditional term is ``get the value of~$s$ if~$\varphi$ holds, or the value of $s'$ if it does not.''
Now, a \emph{(constraint) term} is either a basic term, a conditional term or some (possibly infinite) expression involving basic and conditional terms.
As before, a \emph{(constraint) atom} is an expression involving a (possibly infinite) number of constraint terms.
We denote the set of all constraint terms and atoms by~$\T$ and~$\C$, respectively.

A \emph{formula}~$\varphi$ over~$\C$ is defined as a basic formula above but with~$c \in \C$ being an arbitrary constraint atom rather than a basic one.
Given a constraint term, atom or formula~$\alpha$,
we denote the set of variables occurring in~$\alpha$
by ${\mathit{vars}(\alpha) \subseteq \X}$.

For the semantics,
we start by defining the extended domain as ${\mathcal{D}_{\undefined} \eqdef \mathcal{D} \cup \{\undefined\}}$.
A \emph{valuation} $v$ over $\mathcal{X},\mathcal{D}$ is a function
\mbox{$v:\mathcal{X}\rightarrow\mathcal{D}_{\undefined}$}
where
\mbox{$v(x)=\undefined$}
represents that variable~$x$ is left undefined.
Moreover, if ${X \subseteq \mathcal{X}}$ is a set of variables, valuation
${v|_X: X\rightarrow\mathcal{D}_{\undefined}}$ stands for the projection of $v$ on~$X$.
A valuation $v$ can be alternatively represented as the set
\mbox{$\{ (x,v(x)) \mid x \in \mathcal{X}, v(x)\in\mathcal{D}\}$}
by what no pair $(x,\undefined)$ is in the set.
This representation allows us to use standard set inclusion for comparison.
We thus write $v\subseteq v'$ to mean that
\begin{align*}
\{ (x,v (x)) \mid &x \in \X, v (x)\in\D\}
\\
&\subseteq
\{ (x,v'(x)) \mid x \in \X, v'(x)\in\D\}
\end{align*}
%
The set of all valuations over $\mathcal{X},\mathcal{D}$ is denoted by $\mathcal{V}_{\mathcal{X},\mathcal{D}}$
and $\mathcal{X},\mathcal{D}$ dropped whenever clear from context.

We define the semantics of basic constraint atoms via \emph{denotations},
which are functions
\(
\den{\cdot}:\BC \rightarrow 2^{\mathcal{V}}
\),
mapping each basic constraint atom to a set of valuations.
For instance, each difference constraint like $x-y\leq d$ can be captured by a constraint atom ``$x-y\leq d$''
whose denotation~$\den{\text{``}x-y\leq d\text{''}}$ is given by the expected set:
\begin{align}
\{v\in\mathcal{V}\mid v(x),v(y), d \in \mathbb{Z}, \ v(x)-v(y)\leq d\}
  \label{eq:denotation.difference.constraint}
\end{align}

Satisfaction of constraint atoms involving conditional terms is defined by a previous syntactic \emph{unfolding} of their conditional terms,
using some interpretation to decide the truth values of formulas in conditions.
Formally,
an \emph{interpretation} over $\mathcal{X},\mathcal{D}$ is a pair $\langle h,t \rangle$
of valuations over $\mathcal{X},\mathcal{D}$ such that $h\subseteq t$.
The interpretation is \emph{total} if $h=t$.
With this, we define next two valuation functions for conditional terms,
one following the \emph{vicious cycle principle}~($\gz$)
and another ensuring the \emph{definedness} of conditional terms~($\f$).
%
\begin{definition}\label{def:valuation.function}
  Given an interpretation~$\tuple{h,t}$ and a conditional term $s=(\ctermm{s'}{s''}{\varphi})$,
  we define:
  \begin{align}
    \label{def:eval:term.gz}
    \evalgz{h}{t}(s)
    &=
      \left\{
      \begin{array}{ll}
        s' & \text{if } \langle h,t\rangle\models\varphi\\
        s'' & \text{if } \langle t,t\rangle \hspace{2pt}\not\models\varphi\\
        \undefined & \text{otherwise}
      \end{array}
      \right.
    \\
    \label{def:eval:term.f}
    \evalf{h}{t}(s)
    &=
      \left\{
      \begin{array}{ll}
        s' & \text{if } \langle h,t\rangle\models\varphi\\
        s'' & \text{if } \text{otherwise}
      \end{array}
      \right.
  \end{align}
\end{definition}
%
Note that the valuation functions rely on the satisfaction relation $\models$ defined below.

To illustrate the different behavior of $\gz$ and $\f$,
take the following simple example
\begin{gather}
x=1 \ \leftarrow \ (\ctermm{1}{0}{ x=1 })\geq 0
  \label{eq:gz-f.difference}
\end{gather}
stating that~$x$ must have value~$1$ when the conditional expression~${(\ctermm{1}{0}{ x=1 })\geq 0}$
holds.
We see below that this rule has no stable model under \gzsemantics
while it has a unique one with~${t(x)=1}$ under \fsemantics.
We face here a completely analogous situation to the following standard (non-hybrid) ASP rule with an aggregate:
\begin{gather}
\mathit{holds}_x(1) \ \leftarrow \ \mathit{count}\agg{1:\mathit{holds}_x(1)}\geq 0
  \label{eq:gz-f.difference.ASP}
\end{gather}
where predicate atom $\mathit{holds}_x(1)$ is playing the role of $x=1$ in~\eqref{eq:gz-f.difference}.
Rule~\eqref{eq:gz-f.difference.ASP} has the unique stable model~$\{\mathit{holds}_x(1)\}$ under \citeauthor{ferraris11a}' semantics for aggregates,
which does not comply with \gz,
whereas it has no stable model%
\footnote{~\citeauthor{gelzha19a}'s semantics is defined exclusively for set based aggregates, but lifting it to multi-sets is straightforward.}
under~\citeauthor{gelzha19a}'s semantics,
which satisfies~\gz. 
\begin{observation}\label{obs:eval.total}
Given some total interpretation~$\tuple{t,t}$ and any conditional term $s$,
we have~${\evalgz{t}{t}(s)=\evalf{t}{t}(s)}$.
\end{observation}
Hence, for total interpretations,
we may just write~${\evalcl{t}(s)}$ instead of~${\evalgz{t}{t}(s)}$ and~${\evalf{t}{t}(s)}$.
In our running example,
if~${\tuple{t,t}}$ is a total interpretation such that~${\tuple{t,t} \models (x=1)}$,
then~${\evalcl{t}(\ctermm{1}{0}{ x=1 })}$ is~$1$.
This means that to evaluate whether~$\tuple{t,t}$ satisfies \eqref{eq:gz-f.difference} wrt any of the two semantics,
we replace the conditional expression by the domain element~$1$ and, thus,
evaluate whether~$\tuple{t,t}$ satisfies the basic formula
\begin{gather}
x=1 \ \leftarrow \ 1 \geq 0
  \label{eq:gz-f.difference.undefined}
\end{gather}
which obviously holds since we assumed~${\tuple{t,t} \models (x=1)}$.
For non-total interpretations, the valuation is slightly more involved,
so we will resume our example after introducing the definition of the satisfaction relation.

We permit that different occurrences of conditional expressions are interpreted according to different
valuation functions ($\gz$ or~$\f$).
This can be simply achieved by some syntactic distinction like, for instance, enclosing the expression with $( \cdot )$ for $\gz$ and with $[ \cdot ]$ for $\f$.
This allows us assigning different interpretations to different occurrences of the same expression, e.g., in the formula
\begin{gather}
x=1 \ \leftarrow \ (\ctermm{1}{0}{ x=1 })\geq 0 \ \vee \ \neg[\ctermm{1}{0}{ x=1 }] \geq 0
  \label{eq:gz-f.syntactic}
\end{gather}
In order to abstract from the particular syntax used, we just assume that there exists some \kmapping~$\sem$ that tells us, for each occurrence of a conditional term $s$, which evaluation function must be used, that is, either ${\evals{\sem}{h}{t}(s)=\evalgz{h}{t}(s)}$ or ${\evals{\sem}{h}{t}(s)=\evalf{h}{t}(s)}$.
%

For a constraint atom ${c \in \C}$,
we define~$\evals{\sem}{h}{t}(c)$ as the basic constraint atom that results from replacing each conditional term~$s$ in~$c$
by the basic term~$\evals{\sem}{h}{t}(s)$.
%
\begin{definition}\label{def:satisfaction}
  Given a denotation $\den{\cdot}$,
  a \kmapping~\sem,
  an interpretation $\langle h,t \rangle$ \emph{satisfies} a formula~$\varphi$,
  written $\langle h,t \rangle \modelss \varphi$,
  if 
  \begin{enumerate}
  \item $\langle h,t \rangle \modelss c \text{ if } w\in \den{\evals{\sem}{w}{t}(c)}\text{ for }w\in\{h,t\}$
    \label{item:htc:atom}
  \item $\langle h,t\rangle \modelss \varphi \land \psi \text{ if }  \langle h,t\rangle \modelss \varphi \text{ and }  \langle h,t\rangle \modelss \psi$
  \item $\langle h,t\rangle \modelss \varphi \lor \psi \text{ if }  \langle h,t\rangle \modelss \varphi \text{ or }  \langle h,t\rangle \modelss \psi$
  \item $\langle h,t\rangle \modelss \varphi \rightarrow \psi
    \text{ if }\langle w,t\rangle \not\modelss \varphi \text{ or } \langle w,t\rangle \modelss \psi
    \\\hspace*{70pt}\text{ for }w\in\{h,t\}$
  \end{enumerate}
\end{definition}
%
We say that~$\tuple{h,t}$ is a \mbox{$\sem$-model} of~$\varphi$ when ${\tuple{h,t} \modelss \varphi}$.
In particular, $\gz$- and~$\f$-models are those corresponding to evaluating all conditional terms according to~$\gz$ or~$\f$, respectively.
Furthermore,
we may just write ${\langle h,t \rangle \models \varphi}$ when $\varphi$ is a basic formula or when~$\tuple{h,t}$ is total,
because the valuation function becomes irrelevant in those cases.
Note that this satisfaction relation without subindex is the one used in Definition~\ref{def:valuation.function} for the valuation function.
In the rest of the paper, we assume a fixed underlying denotation for constraint atoms.
If not explicitly stated otherwise,
we also assume a fixed underlying \kmapping.

It is worth noting that Definition~\ref{def:satisfaction} differs from~\cite{cafascwa2020a}
in Condition~\ref{item:htc:atom}, which in our setting corresponds to:
\begin{enumerate}
\item[\ref{item:htc:atom}'.] $\langle h,t \rangle \modelsgz c \text{ if } h\in \den{\evals{\gz}{h}{t}(c)}$
\end{enumerate}
That is, satisfaction of an atom was only checked on the here world $h$ and the \kmapping\ was fixed to $\gz$. 
In fact, the satisfaction relation was not parameterized with $\sem$, since the unique valuation function used was $\gz$.
The following result%
\footnote{An extended version of the paper including all proofs can be found here: \url{https://arxiv.org/abs/2003.04176}}
states that our semantics parameterized with the~$\gz$ mapping actually
corresponds to the semantics we introduced in~\cite{cafascwa2020a}.
%
\begin{proposition}\label{prop:htcc:gzprime}
Let~$\varphi$ be a formula and~$\tuple{h,t}$ be some interpretation.
Then,
${\tuple{h,t} \modelsgz \varphi}$
iff
${\tuple{h,t}}$ is a model of~$\varphi$ according to~\cite{cafascwa2020a}.
\end{proposition}

To illustrate satisfaction of formulas under $\gz$, take again~\eqref{eq:gz-f.difference} and suppose we have some $\tuple{h,t}$ where~${h(x)=\undefined}$ and~${t(x) = 1}$.
Then,~${\tuple{t,t}}$ satisfies~${x=1}$
and, as we saw above, this implies that~$\tuple{t,t}$ satisfies~\eqref{eq:gz-f.difference}.
On the other hand,
we also can see that~${\tuple{h,t}}$ does not satisfy~${x=1}$
and, by definition,
we get: ${\tuple{h,t} \modelsgz (\ctermm{1}{0}{ x=1 })\geq 0}$
iff
both
${\tuple{h,t} \models \undefined \geq 0}$
and
${\tuple{t,t} \models 1 \geq 0}$.
In fact,
in view of Proposition~\ref{prop:htcc:gzprime},
it is enough to check whether
${\tuple{h,t}}$ satisfies ${\undefined \geq 0}$.
That is,
${\tuple{h,t} \models \eqref{eq:gz-f.difference}}$
iff~${\tuple{h,t}}$ satisfies the formula
\begin{gather}
x=1 \ \leftarrow \ (\undefined \geq 0)
  \label{eq:gz-f.difference.undefined}
\end{gather}
which holds because $\tuple{h,t} \not\models (\undefined \geq 0)$.

A \emph{theory} is a set of formulas.
An interpretation~$\tuple{h,t}$ is a \emph{$\sem$-model} of some theory~$\Gamma$,
written ${\tuple{h,t} \modelss \Gamma}$,
when ${\tuple{h,t} \modelss \varphi}$ for every ${\varphi \in \Gamma}$.
A formula $\varphi$ is a \emph{tautology} (wrt some underlying denotation and \kmapping) when~\mbox{$\langle h,t \rangle\modelss\varphi$} for every interpretation~\mbox{$\langle h,t \rangle$}.
Note that, this implies that a basic constraint atom~$c \in \BC$ is tautologous whenever \mbox{$\den{c}=\mathcal{V}$}.
%
\begin{definition}\label{def:equilibrium}
A (total) interpretation $\langle t,t\rangle$ is a \emph{\sequilibrium model} of a theory~$\Gamma$,
if ${\tuple{t,t} \modelss \Gamma}$ and there is no ${h\subset t}$
such that ${\tuple{h,t}\modelss \Gamma}$.
\end{definition}
%
Valuation~$t$ is also called a \emph{$\sem$-stable model} of a set of formulas~$\Gamma$
when~$\tuple{t,t}$ is an \sequilibrium model of~$\Gamma$.
For the case of $\gz$-stable we get the next result.
%
\begin{corollary}\label{cor:htcc:gzprime}
The \gzstable models of any theory
coincide with its stable models according to~\cite{cafascwa2020a}.
\end{corollary}

Following with our example above,
it is easy to see that the interpretation we had, $\tuple{h,t}$ with $t(x)=1$ and $h(x)=\undefined$,
is not a $\gz$-stable model of~\eqref{eq:gz-f.difference}
because
${\tuple{h,t} \modelsgz \eqref{eq:gz-f.difference}}$.
If we consider, instead, the $\f$-semantics, we will see that $t$ is in fact a $\f$-stable model of~\eqref{eq:gz-f.difference}.
This is because no $h' \subset t$ forms a model $\tuple{h',t}$. 
We will prove it for $h'=h$ where $h(x)=\undefined$ and the proof for other interpretations is similar.
First, note that for the $\f$-semantics, Condition~\ref{item:htc:atom} of Definition~\ref{def:satisfaction} always uses the evaluation in both worlds $h$ and $t$.
It does not suffice with using $h$, as happened with $\gz$ (Proposition~\ref{prop:htcc:gzprime}).
Hence,
to satisfy
${\tuple{h,t} \modelsf (\ctermm{1}{0}{ x=1 })\geq 0}$
we need 
both
${\tuple{h,t} \models 0 \geq 0}$
and
${\tuple{t,t} \models 1 \geq 0}$.
As a result,
${\tuple{h,t} \modelsf \eqref{eq:gz-f.difference}}$
iff~${\tuple{h,t}}$ satisfies the formula
\begin{gather}
x=1 \ \leftarrow \ (0 \geq 0) \wedge \neg\neg(1 \geq 0)
  \label{eq:gz-f.difference.zero}
\end{gather}
which does not hold.
Just note that the right hand side is a tautology and that~${\tuple{h,t}}$ does not satisfy~${x=1}$
because ${h(x) = \undefined}$.
Hence,
$t$ is a $\f$-stable model of~\eqref{eq:gz-f.difference}.
The following result generalizes this double negation formalization.
%
\begin{proposition}\label{prop:htcc:evalt}
Let~$\tuple{h,t}$ be an interpretation and~${c \in \C}$ be a constraint atom.

Then,~${\tuple{h,t} \modelss c}$ iff~${\tuple{h,t} \models \evals{\sem}{h}{t}(c) \wedge \neg\neg  \evalcl{t}(c) }$.
\end{proposition}
%

The reason why we need to check both worlds for the \mbox{$\f$-semantics}
is to keep the \emph{persistence property} of \HT
as stated in the following result.
%
\begin{proposition}[Persistence]\label{prop:htcc:persistence}
Let $\langle h,t\rangle$ and $\langle t,t\rangle$ be two interpretations,
and $\varphi$ be a formula.

Then,
\(
\langle h,t\rangle \modelss \varphi$ implies $\langle t,t\rangle \modelss \varphi
\).
\end{proposition}
%
The need for the additional evaluation in $t$ comes from the fact that, under $\f$ valuation, some constraint atoms $c$ may satisfy $h \in \den{\evals{\f}{h}{t}(c)}$ but $t \not\in \den{\evals{\f}{h}{t}(c)}$, and so, if we only used $h$, persistence would be violated.
To illustrate this feature, take the conditional constraint atom
\begin{gather}
(\ctermm{1}{2}{ x=1}) \ \geq \ 2
  \label{eq:f.persistence}
\end{gather}
and, again, interpretation~$\tuple{h,t}$ with ${h(x)=\undefined}$ and~${t(x)=1}$.
Then,
we obtain
\begin{align*}
\evalf{h}{t}(\ctermm{1}{2}{ x=1}) \quad&=\quad 2
\\
\evalcl{t}(\ctermm{1}{2}{ x=1}) \quad&=\quad 1
\end{align*}
because ${\tuple{h,t} \not\models (x=1)}$
and ${\tuple{t,t} \models (x=1)}$.
But then
\begin{align*}
h &\in \den{\evalf{h}{t}(\ctermm{1}{2}{ x=1}) \geq 2}
\\
t &\notin \den{\evalcl{t}(\ctermm{1}{2}{ x=1}) \geq 2}
\end{align*}

The following proposition tells us that some other usual properties of \HT are still valid in this new extension.
Let us introduce some notation first.
Given any \HT\ formula~$\varphi$, let $\varphi[\overline{a}/\overline{\alpha}]$ denote the uniform replacement of atoms $\overline{a}=(a_1,\dots,a_n)$ in~$\varphi$ by \HTC\ formulas $\overline\alpha = (\alpha_1,\dots,\alpha_n)$.
%
\begin{proposition}\label{prop:htcc:properties}
  Let $\langle h,t\rangle$ and $\langle t,t\rangle$ be two interpretations,
  and $\varphi$ be a formula. Then,
  \begin{enumerate}
  \item $\langle h,t\rangle \modelss \varphi\rightarrow\bot$ iff $\langle t,t\rangle \not\modelss \varphi$\ ,
    \label{item:2:prop:htcc:properties}
  \item If $\varphi$ is an \HT\ tautology then $\varphi[\overline{a}/\overline{\alpha}]$ is an \HTC\ tautology.%
    \label{item:3:prop:htcc:properties}
  \end{enumerate}
\end{proposition}
%
As an example of Property~\ref{item:3:prop:htcc:properties} in Proposition~\ref{prop:htcc:properties}, we can conclude, for instance, that
$${( x-\cterm{y}{3}{p} \leq 4 ) \to \neg \neg ( x-\cterm{y}{3}{p} \leq 4 )}$$
is an \HTC\ tautology because we can replace $a$ in the \HT\ tautology $a \to \neg \neg a$ by the \HTC\ formula $( x-\cterm{y}{3}{p} \leq 4 )$.
%
In particular,
the second statement guarantees that all equivalent rewritings in~$\HT$ are also applicable to~$\HTC$.
%
%

\section{Terms and assignments}
As said before, the use of terms as subexpressions will be convenient to derive structural properties of constraint atoms.
We will sometimes refer to a constraint atom using the notation $c[s]$ meaning that
the expression for~$c$ contains some distinguished occurrence of subexpression~$s$.
We further write $c[s/s']$ to represent the syntactic replacement in $c$ of subexpression $s$ by $s'$.
Then, we assume the following syntactic properties:
\begin{enumerate}
\item if ${s \in \T}$ is a term,
  then there are constraint atoms in~$\C$ of the form~${s=s}$ and~${s = d}$ for every domain element~${d \in \D}$,%
  \label{item:term.eval}

\item if ${s \in \T}$ is a term,~$c[s] \in\C$ is a constraint atom
  and ${s' \in \T^e}$,
  then $c[s / s'] \in \C$ is also a constraint atom,%
  \label{item:term.substitution.a}

\item if ${s,s' \in \T}$ are terms such that $s'$ is a subexpression of~$s$
  and~$c[s] \in\C$ is a constraint atom,
  then $c[s / s'] \in \C$.
  \label{item:term.substitution.b}

\end{enumerate}
Intuitively,
Condition~\ref{item:term.eval} states that we always define, at least, equality constraint atoms that allow comparing a term $s$ to any domain element or to itself.

Atom ${s=s}$ is not a tautology: it is satisfied iff $s$ has some value.
For this reason, we will sometimes abbreviate ${s=s}$ as~$\df{s}$ meaning that~$s$ is defined.
Conditions~\ref{item:term.substitution.a} and~\ref{item:term.substitution.b} state that replacement of terms must lead to syntactically valid expressions.
Contrarily to first order logic, we only require that terms can be replaced by some particular class of terms rather than all possible terms.
In particular,
Condition~\ref{item:term.substitution.a}
implies that replacing any term by an elementary term must lead to
syntactically valid expressions.
Condition~\ref{item:term.substitution.b} is similar but for every term that is also a subexpression.
For instance,
if ${x-y}$ and ${x-y+z}$ are terms and
${x-y+z \leq 4}$ is a constraint atom,
then we must allow for forming constraint atoms
${x \leq 4}$ and ${x-y \leq 4}$ and ${\undefined +z \leq 4}$
among others.
Note that, since ${x-y+z}$ is not a subexpression of ${x-y}$,
we do not require
${x-y+z+z \leq 4}$ to be a constraint atom.
%

These intuitions are further formalized by imposing the following semantic properties
for any denotation~$\den{\cdot}$,
basic atom~${c\in\BC}$,
basic term~${s\in\BT}$,
domain element~${d \in \D}$,
variable~${x \in \X}$, and
any pair of valuations~$v,v' \in \mathcal{V}$:
\begin{enumerate}[start=5]
\item if ${v(x)=v'(x)}$ for all ${x \in \mathit{vars}(c)}$ then
\\${v \in \den{c}}$ iff $v' \in \den{c}$.
\label{den:prt:1}

\item   $v \in \den{c}$ and $v \subseteq v'$ imply $v' \in \den{c}$,
\label{den:prt:2}

\item $v \in \den{ c[s/\undefined ] }$ implies
$v \in \den{ c[s] }$
\label{den:prt:3}

\item $\den{ d = d } = \mathcal{V}$,
\label{den:prt:eq.1}


\item $\den{ x = d } = \{ v \in \mathcal{V} \mid v(x) = d \}$,
\label{den:prt:eq.2}

\item $\den{ s = d } \cap \den{ s = d' } = \emptyset$ for any $d' \in \D$ with $d \neq d'$
\label{den:prt:eq.3}

\item if~${v \in \den{ s = s' }}$ for any term~${s' \in \T}$,
then there is some ${d' \in \D}$ such that ${v \in \den{ s = d'}}$ and ${v \in \den{ s' = d'}}$
\label{den:prt:eq.4}

\item if $v \in \den{ s = d }$, then
$v \in \den{c[s]}$ iff $v \in \den{c[s/d]}$,
\label{den:prt:8}

\item if $v \notin \den{ s=s }$ and
$v \in \den{c[s]}$, then $v \in \den{c[s/\undefined]}$.
\label{den:prt:10}

\end{enumerate}
Condition~\ref{den:prt:1} asserts that the denotation of $c$ is fixed by combinations of values for $\vars{c}$,
while other variables may vary freely, consequently becoming irrelevant.
Condition~\ref{den:prt:2} makes constraint atoms behave monotonically.
Condition~\ref{den:prt:3} is the counterpart of Condition~\ref{den:prt:2} for terms.
Intuitively,
it says that, if a constraint does not hold for some term,
then it cannot hold when that term is left undefined.
For instance, if we include a constraint atom $x-\cterm{y}{z}{p} \leq 4$,
then we must allow for forming the three constraint atoms $x-y \leq 4$ and $x-z \leq 4$ and $x-\undefined \leq 4$, too,
and any valuation for the latter must also be a valuation for the former two.
Conditions~\mbox{\ref{den:prt:eq.1}-\ref{den:prt:10}} describe the behavior of equality atoms.
Conditions~\ref{den:prt:8} and~\ref{den:prt:10} respectively tell us that, given some valuation $v$, a term $s$ can always be replaced by its defined value ${v(s)=d}$ or by $\undefined$, if it has no value.
Reflexivity and symmetry of `$=$' can be derived from Conditions~\ref{den:prt:eq.4}-\ref{den:prt:8}, as stated below.
%
\begin{proposition}
\label{prop:htcc:termequality}
Given terms $s$, $s'$ and~$s''$,
the following conditions hold:
\begin{enumerate}
\item if ${v \in \den{ s = s' }}$ and ${v \in \den{ s' = s'' }}$,
then ${v \in \den{ s = s'' }}$.

\item if ${v \in \den{ s = s' }}$,
then ${v \in \den{ s' = s }}$, and

\item $\den{\undefined = s} = \den{s = \undefined} = \den{\undefined = \undefined} = \emptyset$.
	\label{item:3:prop:htcc:termequality}
\end{enumerate}
\end{proposition}
%
Condition~\ref{item:3:prop:htcc:termequality} implies that~`$=$' is not reflexive.
Also,
when ${v(x) = \undefined}$, atom ${x=x}$ is false, i.e., ${v \notin \den{ x = x }}$.
%
%
%
The following interesting properties for~$\df{s}$ can also be derived.

\begin{observation}
The following conditions hold:
\begin{enumerate}
\item $\den{ \df{s} } = \bigcup_{d \in \D} \den{s = d}$ for every term~$s \in \T$,
\item $\den{ \df{d} } = \V$ for every domain element~$d \in \D$,
\item ${\den{ \df{x } } = \{ v \in \V \mid v(x) \neq \undefined \}}$ for every~$x \in \X$,
\item $\den{ \df{\undefined} } = \emptyset$.
\end{enumerate}
\end{observation}

These conditions together imply that
constraint terms behave similar to first order terms.
That is,
we can define a recursive function for valuation of terms and subterms:

\begin{definition}[Term valuation]
\label{def:term.valuation}
Given
a valuation~${v \in \V}$,
an interpretation~$\tuple{h,t}$
and
a term~${s \in \T}$,
we define
${v_{\tuple{h,t}}^\sem : \T \longrightarrow \Du}$ as the following function:
\begin{gather*}
v_{\tuple{h,t}}^\sem(s) \ \eqdef \ \begin{cases}
d &\text{if } v \in \den{ \evals{\sem}{h}{t}(s) = d } \text{ with } d \in \D,
\\
\undefined &\text{otherwise}
\end{cases}
\end{gather*}
where
${\evals{\sem}{h}{t}(s)}$ denotes the basic term that results from replacing each conditional term~$s'$ in $s$ by
$\evals{\sem}{h}{t}(s')$.
\end{definition}
%
%
For a constraint atom~${c \in \C}$,
we denote by~$v{_{\tuple{h,t}}^\sem(c)}$ the constraint atom obtained by replacing each occurrence of term~$s$ in $c$ by~${v_{\tuple{h,t}}^\sem(s) \in \Du}$.
Note that $v_{\tuple{h,t}}^\sem(c)$ for constraint atom $c$ becomes a syntactic transformation.
For example, take as $c$ the following constraint atom:
$$(\ctermm{1}{0}{ x=1})-y\ \geq \ 0$$
which is a slight elaboration of the atom in the body of \eqref{eq:gz-f.difference}.
Suppose that we define $s=(\ctermm{1}{0}{ x=1})$ as a term (remember $x$, $y$, $1$ and $2$ are also elementary terms).
Assume also that $\sem$ applied to $s$ in this case selects $\f$ and take the interpretation $\tuple{h,t}$ where $h(x)=\undefined$, $t(x)=1$ as in previous examples, adding now $h(y)=t(y)=1$.
Then, using $\f$, the conditional term is replaced by $0$ in $h$ and by $1$ in $t$.
Therefore $v_{\tuple{h,t}}^\sem(s)=0$ and $v_{\tuple{t,t}}^\sem(s)=1$.
Given that the value of $y$ is fixed to 1, we get the basic atoms $v_{\tuple{h,t}}^\sem(c)=0-1\geq 0$ and $v_{\tuple{h,t}}^\sem(c)=1-1\geq 0$.
%
\begin{proposition}
\label{prop:htcc:eval.valuation}
Given a valuation~$v$, an interpretation~$\tuple{h,t}$, a \kmapping\ $\sem$,
and an atom~$c$,
the following two conditions are equivalent:
\begin{enumerate}
\item $v \in \den{\evals{\sem}{h}{t}(c)}$
\item $v \in \den{v_{\tuple{h,t}}^\sem(c)}$
\end{enumerate}
\end{proposition}
In other words,
we can safely use the term valuation~$v_{\tuple{h,t}}^\sem$
to replace every term for its value in the valuation~$v$ wrt the interpreation~$\tuple{h,t}$.
Note that, in practice, we chose $v$ to be either~$h$ or~$t$.
When $s$ is a basic term or $\tuple{h,t}$ is total, the value returned by $v_{\tuple{h,t}}^\sem(s)$ does not depend on $\sem, h$ or $t$.
For this reason, we just write $v(s)$ in those cases.

Finally,
we can establish some relation between the \gz- and \f-semantics
based on how they evaluate constraint terms.

\begin{proposition}\label{prop:term:persistence}
Any term~$s$ and interpretation~$\tuple{h,t}$ satisfy that
${h_{\tuple{h,t}}^\gz(s)\neq \undefined}$
implies
${h_{\tuple{h,t}}^\gz(s)=h_{\tuple{h,t}}^\f(s)=t(s)}$.
\end{proposition}
In other words, if the \gzsemantics assigns some value to term $s$ in~$h$,
this value is also preserved in $t$.
Moreover,
when this is the case the \gz- and \f-semantics coincide.
On the other hand,
this preservation property is not satisfied by the \f-semantics, as we discussed in the example with the conditional atom in~\eqref{eq:f.persistence}.


As mentioned in the introduction,
the main motivation to introduce constraint terms is to permit a uniform treatment of different constructs.
This is especially relevant for assignments~\cite{cakaossc16a}.
Intuitively,
an assignment of the form~${x := s}$ is a \emph{directional} construct meaning that variable~$x$ takes the value of term~$s$.

\begin{definition}[Assignment]
Given a variable~$x\in\X$ and a term~$s\in\T$
an \emph{assignment} is an expression of the form~${x := s}$ and stands for the formula
\begin{align}
x = s \ \leftarrow \ \df{ s }
  \label{f:aggs.assig}
\end{align}
\end{definition}
Recall that, in~\cite{cakaossc16a},
assignments were constructs where~$s$ could only be a linear expression.
The introduction of terms
allows us to generalize the use of assignments to arbitrary terms which,
 as we can see in following sections, includes both linear expressions and aggregates.
The following result provides further intuition relating assignments with grounding in ASP.
%
\begin{theorem}\label{thm:grounding.agg.assigments}
Consider a formula of the form
\begin{align}
x := s \ \leftarrow \ \varphi
  \label{eq:1:thm:grounding.agg.assigments}
\end{align}
where~${x \in \X}$ is a variable,~${s\in\T}$ a constraint term and~$\varphi$ a (sub)formula.
Let~$\Gamma$ collect the set of formulas:
\begin{align}
x = d \ \leftarrow \ \varphi \wedge s = d
  \label{eq:2:thm:grounding.agg.assigments}
\end{align}
for every element~${d \in \D}$ in the domain.
Then,~$\Gamma$ and~\eqref{eq:1:thm:grounding.agg.assigments} have the same $\sem$-models.
\end{theorem}
%
In other words,
an assignment on the consequent of an implication
stands for (the possibly infinite grounding of) the first order formula~${\forall Y\, (x=Y \leftarrow \varphi \wedge s = Y)}$.
Of course,
the advantage of assignments consists in the possibility of delegating their evaluation to specialized constraint solvers.
For this, such solvers only need to be able to deal with equality constraints.
This also implies that grounding is not necessary.
Note that, if $\D$ is infinite, then so is $\Gamma$.

\section{Aggregates as constraint atoms} 
\label{sec:aggregates}

Aggregates are expressions that represent a function that groups together a collection of expressions and produces a single value as output.
For instance,
the expression
\begin{gather}
\mathit{sum}\agg{ \, \mathit{tax}(P) : \mathit{lives}(P,R) \, }
  \label{eq:aggregate.tax}
\end{gather}
shown in the introduction
sums the tax revenue of all persons~$P$ that live in some region~$R$.
In this section, we restrict ourselves to ground atoms,
that is, atoms that may contain constraint variables
but no logical variables like~$P$ and~$R$.
We assume that aggregate atoms with logical variables are a shorthand for their (possibly infinite) ground instantiation.
For instance,~\eqref{eq:aggregate.tax} is a shorthand for the infinite expression of the form
$\mathit{sum}\agg{ \alpha_1, \alpha_2, \dotsc }$
where each~$\alpha_1, \alpha_2, \dotsc$ is a sequence containing a conditional term of the form~${\mathit{tax}(p) : \mathit{lives}(p,r)}$
for each pair of domain elements~$p$ and~$r$.
Intuitively,
the variable~${\mathit{lives}(p,r)}$ is true when the person~$p$ lives in the region~$r$
and the variable~${\mathit{tax}(p)}$ is assigned the tax revenue of person~$p$.
If~$p$ is not a person,~${\mathit{tax}(p)}$ is undefined, that is, its assigned value is~$\undefined$.
As a simpler example, we have the following expression
\begin{gather}
\htagg{sum}{\ 1\!:\!p,\ 1\!:\!q ,\ 2\!:\!r \ }{\geq}{2}
  \label{eq:ex.agg.sum}
\end{gather}
which holds if either $r$ holds (regardless of the other variables) or both $p$ and~$q$~hold, otherwise.
More generally,
we allow applying aggregates not only to numerical constants,
but also to expressions involving constraint variables.
For instance,
\begin{gather}
\htagg{sum}{\ x\!:\!p,\ y+z\!:\!q  \ }{\geq}{2}
  \label{eq:ex.agg.sum.with.var}
\end{gather}
holds whenever any of the following conditions hold:
\begin{itemize}
\item $x \geq 2$ and only~$p$ holds,
\item $y+z \geq 2$ and only~$q$ holds, or
\item $x+y+z \geq 2$ and both $p$ and~$q$ hold.
\end{itemize}
We also allow aggregate operations that rely on the order of the elements in the collection.
For instance,
the aggregate
\begin{gather}
\appendblank\langle\text{``En'', ``un'', ``lugar'', ``de" , ``la'', ``Mancha''} \rangle=x
  \label{eq:ex.agg.string}
\end{gather}
expresses that $x$ is the string resulting from concatenating all strings occurring between the brackets.
That is,
it is only satisfied when the value assigned to~$x$ is the string ``En un lugar de la Mancha''.

Formally,
an \emph{aggregate term} is an expression of the form
\begin{gather}
\op\langle s_1, s_2, \dotsc \rangle
  \label{eq:agg.inf}
\end{gather}
where~$\op$ is an \emph{operation symbol}
and each~$s_i\in \T$ is a term.
We say that a term is \emph{aggregate-free} if it contains no aggregate terms and, in the following,
we assume that each~$s_i$ in~\eqref{eq:agg.inf} is aggregate-free.
A \emph{basic aggregate term} is an expression of the form of~\eqref{eq:agg.inf} where each~$s_i$ is a basic term.
We reserve the notation~$\agg{\dotsc}$ for aggregates whose operation is multi-set based, like~$\mathit{sum}$, and use~$\langle\dotsc \rangle$ in general.

An infinite sequence $\theta$ of domain elements $\tuple{d_0,d_1, \dots}$ can be defined as a mapping $\theta:(\mathbb{N}^+ \to \D)$ so that $\theta(i)=d_i$ for all $i\geq 0$.
Notice that $\theta$ may contain repeated occurrences of the same domain value.
We sometimes denote a finite prefix $\theta'=\tuple{d'_0,\dots,d'_{n-1}}$ of length $n\geq 0$ and use the concatenation $\theta' \cdot \theta$ to yield an infinite sequence defined as expected $\tuple{d'_0,\dots,d'_{n-1},d_0,d_1,\dots}$.

Given each aggregate term like~\eqref{eq:agg.inf}, we assume there exists an associated fixed operation
\(
{\hat{\op}:(\mathbb{N}^+ \to \D) \to \Du}
\)
assigning a domain value $d \in \D$ (or $\undefined$) to any infinite sequence of domain values.
As an example, in the case of the $\mathit{sum}$ aggregate we get $\hat{sum} \langle d_1, d_2, \dots \rangle = \sum_{i\geq 0} d_i$ as expected.
Depending on the domain and the operator, we may sometimes obtain $\undefined$ as a result.
For instance, if $\D$ are the natural numbers and we sum an infinite sequence of $1$'s, the result of $\hat{sum}\langle 1,1,1,\dots \rangle$ is not a natural number and the sum would be undefined $\undefined$.
We say that some~$0_\op \in \D$ is a \emph{neutral element} for $\op$ if for all infinite sequence $\theta$ and any finite prefix $\theta'$ we have ${\hat{\op}(\theta' \cdot 0_\op \cdot \theta) = \hat{\op}(\theta' \cdot \theta)}$. 
Without loss of generality, we restrict ourselves to operations~$\op$ that have a neutral element.
Otherwise, we can always build an equivalent function with neutral element by adding a new element to the domain.

\begin{definition}[Evaluation of a basic aggregate term]
We define the \emph{evaluation} $v(A)$ of a basic aggregate term~$A$ like~\eqref{eq:agg.inf} with respect to a valuation $v$ as
\begin{gather}
\label{eq:aggregate.valuation}
v(A) \ \eqdef \ \begin{cases}
\hat{\op}(\theta_A) &\text{if } v(s_i) \neq \undefined \text{ for all } i \geq 1
\\
\undefined &\text{otherwise}
\end{cases}
\end{gather}
where
$
{\theta_A: \mathbb{N}^+ \to \D}
$
is a function mapping each positive integer~${i \in \mathbb{N}^+}$ to the value~$v(s_i)$.
\end{definition}

An \emph{aggregate atom} (or \emph{aggregate} for short) is an expression of the form~${A \prec s_0}$
where~$A$ is an aggregate term,~$\prec$ is a relation symbol and~$s_0$ is a basic term.
We associate the symbol~$\prec$ with some relation
\(
{\hat{\prec} \subseteq \D \times \D}
\)
among elements of the domain.
The denotation of a basic aggregate atom is then defined as
\begin{align*}
  \den{A \prec s_0}
  \ \eqdef \  \{v\in\mathcal{V}\mid v(A) \mathrel{\hat{\prec}} v(s_0) \}
\end{align*}
In particular, note that $\prec$ can be the equality symbol.
The semantics of conditional aggregates follows directly from the evaluation of conditions introduced in the previous section.

The following result shows how the evaluation of terms introduced in Definition~\ref{def:term.valuation}  applies to the particular case of aggregate terms.
%
\begin{proposition}[Evaluation of an aggregate term]
\label{prop:htcc:aggtermeval}
We define the \emph{evaluation} of an aggregate term~$A$ possibly containing conditional terms,
with respect to some valuation~$v \in \V$, some interpretation~$\tuple{h,t}$
and a \kmapping~$\sem$,
as
\begin{gather*}
v_{\tuple{h,t}}^\sem(A) \ \eqdef \ \begin{cases}
\hat{\op}(\theta_{A,\tuple{h,t}}^\sem) &\text{if } v_{\tuple{h,t}}^\sem(s_i) \neq \undefined \text{ for all } i \geq 1
\\
\undefined &\text{otherwise}
\end{cases}
\end{gather*}
where
$
{\theta_{A,\tuple{h,t}}^\sem: \mathbb{N}^+ \to \D}
$
is a function mapping each positive integer~${i \in \mathbb{N}^+}$ to the value~$v_{\tuple{h,t}}^\sem(s_i)$.
\end{proposition}
\begin{corollary}
Given an aggregate term~$A$ (possibly with conditional terms),
a valuation~$v \in \V$, some interpretation~$\tuple{h,t}$
and a \kmapping~$\sem$, we have:

${\tuple{h,t} \models A \prec s_0}$
iff ${v_{\tuple{v,t}}^\sem(A) \mathrel{\hat{\prec}} v_{\tuple{v,t}}^\sem(s_0)}$
for~${v \in \{h,t\}}$. 
\end{corollary}

Using neutral elements,
we can consider finite aggregates as abbreviations for infinite ones.
That is, a finite constraint term of the form
\begin{gather}\label{eq:cfree.aggreagate}
  \op\langle s_1, s_2, \dots, s_n \rangle
\end{gather}
is an abbreviation for the infinite term
\begin{gather}\label{eq:cfree.aggreagate}
  \op\langle s_1, s_2, \dots, s_n, 0_\op, 0_\op, \dotsc \rangle
\end{gather}
Treating finite aggregates as an abbreviation allows us to deal with a unique construct for any number of elements and, thus,
ensure that aggregate terms with different number of elements are treated in an uniform way.

We also adopt some further conventions for multi-set based aggregates
that reflect the syntax of ASP solvers.
In particular,
a \emph{multi-set aggregate term}
is an expression of the form
\begin{align}\label{eq:multiset.aggreagate}
  \op\agg{ \tau_1, \tau_2, \dotsc}
\end{align}
where each $\tau_i$ is either a basic term
or an expression of the form $s_i' : \varphi_i$
with $s_i'$ a basic term and~$\varphi_i$
a basic formula.
Such an expression is understood as an abbreviation for an aggregate term of the form of~\eqref{eq:agg.inf}
where each~$s_i$ is as follows:
\begin{enumerate}
\item $s_i = \cterm{\tau_i}{0_\op}{\df{\tau_i}}$ if $\tau_i$ is a basic term, and
\item $s_i = \cterm{s_i'}{0_\op}{\df{s_i'} \wedge \varphi_i }$ otherwise.
\end{enumerate}

This allows us to capture the behavior of modern ASP solvers.
For instance, the solver \clingo\ removes elements that are undefined from the \texttt{sum} aggregate and a return value can still be obtained.
Now,
the semantics of~\eqref{eq:ex.agg.sum} and~\eqref{eq:ex.agg.sum.with.var} can be formalized by defining the following function
\begin{gather}
\label{eq:sum.sem.simple}
  \hat{sum}(\theta)
  \ \ \eqdef \ \
\sum \{ \ \theta(i)  \mid i \in\mathbb{N}^+ \text{ and } \theta(i)\in\mathbb{Z} \ \}
\end{gather}
where~${\theta : \mathbb{N}^+ \to \D}$ is a family of domain elements.
For~$\leq$, we take the usual meaning.
Obviously,
the neutral element of~$\mathit{sum}$ is ${0_{\mathit{sum}} = 0}$.
Note that combining this definition with~\eqref{eq:aggregate.valuation},
we get that the sum of an aggregate term is undefined if any of its elements is undefined,
otherwise, we get the sum of all integers in the sequence.

Let us illustrate the behavior of aggregates in our setting taking~\eqref{eq:ex.agg.sum}
as an example.
Note that, following our convention,~\eqref{eq:ex.agg.sum} is a short hand for the atom%
\footnote{We dropped the tautologies $\df{1}$ and $\df{2}$.}
\begin{gather}
\mathit{sum}\tuple{\, \cterm{1}{0}{p},\, \cterm{1}{0}{q} ,\, \cterm{2}{0}{r},\, 0 ,\, 0, \dotsc \,}{\ \geq\ }{2}
  \label{eq:ex.agg.sum.expanded}
\end{gather}
We see that
if $p,q,r$ hold in some interpretation,
then the left hand side of the inequality evaluates to~${\sum\{1,1,2\} = 4}$ and, the inequality is satisfied.
On the other hand,
if only $p,q$ hold,
we get $\sum\{1,1\} = 2$
and the inequality is not satisfied.
As another example,
while evaluating the aggregate term
\begin{gather}
{\mathit{sum}\{ 2,\, 5,\, \text{``hello world''},\, 7 \}}
  \label{eq:sum.hello}
\end{gather}
the string ``hello world'' is ignored and the result is just~$14$.

Beyond arithmetic aggregates,
we may also have expressions like~\eqref{eq:ex.agg.string}, which deal with strings.
We define $0_\appendblank$ as the empty string
and
$\hat{\appendblank}(\theta)$ as the string $\theta(1)\textvisiblespace \theta(2)\textvisiblespace \dots$
resulting of concatenating all strings in~$\theta$.
%

\section{Aggregates as conditional linear constraints}
\label{sec:linconst}

One important difference between the understanding of aggregates used in this paper and the one studied in~\cite{cafascwa2020a}
is that the latter directly considers aggregates as abbreviations for conditional linear constraints.
This viewpoint is interesting because it allows the use of off-the-shelf CASP solvers to compute aggregates with constraint variables.
On the downside, this approach has two drawbacks.
First,
it is quite different from the usual definition of aggregates in the ASP literature,
which makes it difficult to relate to existing approaches in standard (non-constraint) ASP.
Second,
it is more restrictive as it only permits a particular class of aggregates,
namely those using the operation functions~\texttt{sum}, \texttt{count}, \texttt{max} and \texttt{min}.

The definition we provide in the previous section solves these two issues,
but leaves us with the question whether we can use off-the-shelf CASP solver to compute aggregates.
In this section,
we show that it is possible to translate~\texttt{sum} aggregates into
conditional linear constraints.
Thus, affirmatively answering the above question for the \mbox{\gz-semantics}.
In the next section,
we extend this result to an interesting class of theories under the \fsemantics.
Aggregates with operations~\texttt{count}, \texttt{max} and \texttt{min}
can be mapped to~\texttt{sum} ones~\cite{alfage15a}.

We start by reviewing the definition of
\emph{conditional linear constraints} from~\cite{cafascwa2020a},
but incorporating our notion of term.
A \emph{product term} is either an integer~${d \in \Z}$, a variable~${x \in \X}$ or an expression of the form~${d \cdot x}$
where~${d \in \Z}$ is a domain element and~${x \in \X}$ is a variable.
A \emph{finite basic linear term}
is either a product term or an expression of the form~${s_1 + \dotsc + s_n}$ where each~$s_i$ is a product term.
A \emph{linear term}
is an expression of~${s_1 + s_2 + \dotsc}$ where each~$s_i$ is
either a finite basic linear term or a conditional term of the form~$\cterm{s_i'}{s_i''}{\varphi_i}$
with~$s_i'$ and~$s_i''$ finite basic linear terms and~$\varphi_i$ a basic formula.
%
%
A \emph{linear constraint} is a comparison of the forms \mbox{$\alpha \leq \beta$},
\mbox{$\alpha < \beta$},
\mbox{$\alpha = \beta$} or \mbox{$\alpha \neq \beta$} for linear terms~$\alpha$ and~$\beta$.
As usual,
we write~${\alpha \geq \beta}$ and~${\alpha > \beta}$ to stand for ~${\beta \leq \alpha}$ and~${\beta < \alpha}$,
respectively.

We adopt some usual abbreviations.
We directly replace the~`$+$' symbol by (binary) `$-$' for negative constants
and, when clear from the context, omit the `$\cdot$' symbol and parentheses.
We do not remove parentheses around conditional expressions.
As an example,  the expression ${-x + \ \cterm{3y}{2y}{\varphi} \ - 2 z}$ stands for
$(-1) \cdot x + \cterm{3 \cdot y}{2 \cdot y}{\varphi} + (-2) \cdot z$.
Other abbreviations must be handled with care.
In particular, we neither remove products of form $0 \cdot x$ nor replace them by $0$ (this is because~$x$ may be undefined, making the product undefined).

In the rest of the paper,
we assume that all integers are part of the domain, that is,~${\Z \subseteq \D}$.
Given a valuation~$v$,
the semantics of basic linear constraints is defined inductively.
\begin{align*}
v(d \cdot x) &\ \eqdef\  \begin{cases}
\mathrlap{d \cdot v(x)}\hphantom{\sum_{i \geq 1} v(s_i)} &\text{if } v(x) \in \Z
\\
\undefined &\text{otherwise}
\end{cases}
\\
v(s_1 + s_2 + \dotsc) &\ \eqdef\  \begin{cases}
d  &\text{if } \forall i \geq 1, \ v(s_i) \in \Z\\
& \text{and } \sum_{i \geq 1} v(s_i)=d\in\D
\\
\undefined &\text{otherwise}
\end{cases}
\end{align*}
The denotation of a basic linear constraint ${\alpha \prec \beta}$ is given by
\begin{gather*}
\den{\alpha \prec \beta} \ \eqdef \  \{\val \mid \val(\alpha), \val(\beta)\in \mathbb{Z}, \val(\alpha) \prec \val(\beta)\}
\end{gather*}
with~$\prec$ a relation symbol among~$\leq$, $<$, $=$ and~$\neq$.
In particular,
given a linear constraint of the form~${\alpha \leq d}$ with
${\alpha = d_1 \cdot x_1 + \dots + d_n \cdot x_n}$,
we have
${v \in \den{\alpha \leq d}}$
iff
${v(x_i) \neq \undefined}$ for all~${1 \leq i \leq n}$
and
${d \geq \sum_{1\leq i \leq n} d_i \cdot \val(x_i)}$.

The semantics of conditional linear constraints is immediately obtained by applying the corresponding evaluation functions.
The following results show how the valuation function~$v_{\tuple{h,t}}^\sem$ applies to (conditional) linear terms;
and how to use this for evaluating (conditional) linear constraints.
%
\begin{proposition}[Linear term evaluation]
\label{prop:htcc:lintermeval}
Let ${v \in \V}$ be a valuation,
$\tuple{h,t}$ be an interpretation
and
${\alpha=s_1 + s_2 + \dots}$ be a linear term (possibly containing conditional terms).
Then,
\begin{gather*}
v_{\tuple{h,t}}^\sem(\alpha) \ = \ \begin{cases}
d  &\text{if } \forall i \geq 1, \ v_{\tuple{h,t}}^\sem(s_i) \in \Z
\\
& \text{and } \sum_{i \geq 1} v_{\tuple{h,t}}^\sem(s_i) = d \in \D\\
\undefined &\text{otherwise}
\end{cases}
\end{gather*}
\end{proposition}
\begin{corollary}
\label{cor:htcc:lintermeval}
Given an interpretation~$\tuple{h,t}$
and a linear constraint~${\alpha\prec\beta}$ (possibly containing conditional~terms), we get:
${\tuple{h,t} \modelss \alpha\!\prec\!\beta}$
iff ${v_{\tuple{v,t}}^\sem(\alpha) \prec v_{\tuple{v,t}}^\sem(\beta)}$ for both~${v \in \{h,t\}}$.
\end{corollary}
%
The following result shows some interesting equivalences.
%
\begin{proposition}
\label{prop:htcc:linear.decomposition}
Given an interpretation $\tuple{h,t}$
and linear terms~$\alpha$ and~$\beta$
the following equivalences hold:
\begin{enumerate}
\item $\tuple{h,t}\modelss \alpha = \beta$ iff $\tuple{h,t}\modelss \alpha \leq \beta \land \alpha \geq \beta$\label{prop:htcc:linear.decomposition:1},
\item $\tuple{h,t}\modelss \alpha < \beta$ iff $\tuple{h,t}\modelss \alpha\leq\beta \land \alpha \neq \beta$\label{prop:htcc:linear.decomposition:2},
\item $\tuple{h,t}\modelsgz \alpha < \beta$ iff $\tuple{h,t}\modelsgz \alpha\leq\beta \land \neg(\alpha \geq \beta)$\label{prop:htcc:linear.decomposition:3},
\item $\tuple{h,t}\modelsgz \alpha\neq\beta$ iff $\tuple{h,t}\modelsgz \alpha<\beta \lor \alpha>\beta$\label{prop:htcc:linear.decomposition:4}.
\end{enumerate}
\end{proposition}
%
We see that with the \mbox{$\gz$-semantics},
we can define all arithmetic relations in terms of~$\leq$,
while we need~$\leq$ and~$\neq$ for the \mbox{$\f$-semantics}.
To see that the third equivalence does not hold for the \mbox{$\f$-semantics},
take
the interpretation~$\tuple{h,t}$ with~${h(x)=\undefined}$ and ${t(x)=1}$
and the atom
\begin{gather}
\cterm{0}{1}{x=1} \ < \ 1
  \label{ex:linconst:strict.leq}
\end{gather}
Then,
with~$\alpha$ being the linear term on the left hand side of \eqref{ex:linconst:strict.leq},
we get that~${\tuple{h,t} \modelsf (\alpha \leq 1) \wedge \neg(\alpha \geq 1)}$ holds
despite of~${\tuple{h,t} \not\modelsf (\alpha < 1)}$.
Similarly,
for
the last equivalence
take the same interpretation~$\tuple{h,t}$
and the constraint
\begin{gather}
\cterm{0}{1}{x=1} \ \neq \ \cterm{1}{0}{x=1}\label{ex:linconst:uneqf}
\end{gather}
Then,
with~$\alpha$ and $\beta$ being the linear terms on the left and right hand side of \eqref{ex:linconst:uneqf}, respectively,
we get~${\tuple{h,t} \modelsf (\alpha \neq \beta)}$
despite of~${\tuple{h,t} \not\modelsf (\alpha < \beta) \vee (\alpha > \beta)}$.

Let us now show how~$\mathit{sum}$ aggregates can be translated into conditional linear constraints.
First, we introduce a new constraint atom~$\isint{s}$
whose intuitive meaning is that term~$s$ is an integer and whose
denotation is given as follows:
\begin{gather*}
\den{ \isint{s} } \quad \eqdef \quad \bigcup \big\{ \, \den{ s = d} \mid d \in \Z \,\big\}
\end{gather*}

\begin{definition}[Aggregate to linear term]
Given an aggregate term~$A$ of the form~\eqref{eq:multiset.aggreagate}
with ${\op = \mathit{sum}}$,
we associate the linear term ${\pi(A) \eqdef \pi(\tau_1) + \pi(\tau_2) + \dotsc}$
where~$\pi(\tau_i)$ is defined as follows:
\begin{enumerate}
\item $\pi(\tau_i) \eqdef \cterm{\tau_i}{0}{\isint{\tau_i}}$ if $\tau_i$ is a finite basic linear term,
\item $\pi(\tau_i) \eqdef \cterm{s_i}{0}{\isint{s_i} \wedge \varphi_i}$ if $\tau_i$ is of the form ${s_i : \varphi_i}$.
\end{enumerate}
For a theory~${\Gamma}$,
we define~${\pi(\Gamma)}$ as the result of recursively replacing each aggregate term~$A$ by~$\pi(A)$ in~$\Gamma$.
\end{definition}

Furthermore, for a \kmapping{} $\sem$,
we define the \kmapping{} $\pi(\sem)$ given as follows:
\begin{enumerate}
\item ${\pi(\sem)(s)=\sem(s)}$ for every occurrence of conditional term~$s$ not occurring in any aggregate term~$A$,
\item $\pi(\sem)(\pi(s)) = \sem(s)$  for every occurrence of conditional term~$s$ occurring in some aggregate term~$A$.
\end{enumerate}

\begin{theorem}\label{thm:htcc:linconstagg}
For any theory~$\Gamma$,
the $\sem$-(stable) models of~$\Gamma$
and $\pi(\sem)$-(stable) models of~$\pi(\Gamma)$ coincide.
\end{theorem}
This means that we can use the techniques developed in~\cite{cafascwa2020a}
to compute the stable models of theories with aggregate under the \mbox{\gz-semantics}.

For instance,~\eqref{eq:ex.agg.sum.with.var} becomes the linear constraint
\begin{gather*}
\cterm{x}{0}{\isint{x}\land p} + \cterm{z+y}{0}{\isint{z+y}\land q} \geq 2
\end{gather*}
which holds under the same conditions as~\eqref{eq:ex.agg.sum.with.var} does.
As a further example,
the aggregate term~\eqref{eq:sum.hello}
is translated as the linear constraint~${\pi(2) \!+\! \pi(5) \!+\! \pi( \text{``hello world''}) \!+\! \pi(7)}$.
For any integer~$n$,
we get that~${\pi(n) = \cterm{n}{0}{\isint{n}}}$ is simply equivalent to~$n$
because~$\isint{n}$ is a tautology.
On the other hand,
$\pi( \text{``hello world''})$ is equivalent to~$0$
because $\isint{\text{``hello world''}}$ is a contradiction.
Hence, we get
$\pi(2) + \pi(5) + \pi( \text{``hello world''}) + \pi(7) = 2 + 5 + 7 = 14$.
%

\section{Logic programs}
\label{sec:lp}

We focus now on a restricted syntax corresponding to logic programs
and show that, for stratified occurrences of conditional terms,
we can safely exchange $\gz$ and \fsemantics.
A \emph{literal} is either a constraint atom~$c \in \C$
or the negation~$\neg c$ of one.
A \emph{rule} is a formula of the form $H \leftarrow B$
where $H$ is a either an assignment or a disjunction of literals
and $B$ is a conjunction of literals
called the rule's \emph{head} and \emph{body}, respectively.
A theory consisting exclusively of rules is called a \emph{(logic) program}.
Further, we adopt the following conventions.
For any rule~$r$ of form~${H \leftarrow B}$
we let $\Head(r)$ and $\Body(r)$ stand for the set of all literals occurring in~$H$ and~$B$, respectively.
If $H$ is an assignment~${x:=s}$,
we assume that~$B$ contains additionally~$\df{s}$.
We denote the set of positive and negative literals in $\Head(r)$
by ${\Headp(r) \eqdef \Head(r) \cap \C}$ and ${\Headn(r) \eqdef \Head(r) \setminus \Headp(r)}$.
%
We also define
${\varsp{c} \eqdef \{x\}}$ if $c$ is an assignment~${x := s}$
and
${\varsp{c} \eqdef \vars{c}}$ if $c$ is a constraint atom.
Intuitively,
${\varsp{c}}$ designates all variables assigned by atom~$c$\,:
Only the assigned variable is defined by an assignment
and all variables in a constraint atom.
%
\begin{definition}[Conditional term stratification]
\label{def:cond.stratification}
A program $\Pi$ is said to be \emph{stratified on (an occurrence of) a conditional term~$s=\cterm{s'}{s''}{\varphi}$},
if there is a level mapping \mbox{$\levels: \mathcal{X} \longrightarrow \mathbb{N}$}
satisfying the following conditions for every rule~${r \in \Pi}$:
\begin{enumerate}
\item ${\level{x} \geq \level{y}}$ for all variables
${x \in \varsp{\Headp(r)}}$ and ${y \in \vars{\Headn(r) \cup \Body(r)}}$,
\label{item:1:def:cond.stratification}

\item ${\level{x} = \level{y}}$ for all variables ${x,y \in \varsp{\Headp(r)}}$
\end{enumerate}
plus the following condition for the rule~$r$ where~$s$ occurs
\begin{enumerate}
\item  ${\level{x} > \level{y}}$ for all ${x \in \varsp{\Headp(r)}}$ and ${y \in \var(\varphi)}$.
\end{enumerate}
A program~$\Pi$ is \emph{stratified}
if it is stratified on all occurrences of conditional terms not occurring in the scope of negation.
\end{definition}
Given \kmappings~$\sem$,~$\sem'$ and a distinguished occurrence of some conditional term~$s$,
we denote by~$\sem[s \mapsfrom \sem' ]$ the \kmapping obtained from~$\sem$ by replacing the result assigned to~$s$ by the one that~$\sem'$ assigns to it.
\begin{theorem}{\label{thm:gz-f.correspondence}}
Let $\Pi$ be a program stratified on
some occurrence of a conditional term~$s$
and $\sem$ and~$\sem'$ be two \kmappings.
Let~$\sem'' = \sem[s \mapsfrom \sem']$.
Then, the $\sem$-stable models and the $\sem''$-stable models of~$\Pi$ coincide.
\end{theorem}
%
\begin{theorem}{\label{thm:gz-f.correspondence2}}
For a stratified program,
its $\sem$- and $\sem'$-stable models coincide
for any pair of \kmappings~$\sem$ and~$\sem'$.
\end{theorem}
%
This means that,
for stratified programs, we can use the translation from~\cite{cafascwa2020a} to rely on off-the-shelf CASP solvers to compute not only the $\gz$-stable models but also the $\f$-stable models.
Furthermore, as our concept of stratification and the translation pertain to distinguished occurrences of conditional terms,
it is possible to partially translate non-stratified programs for stratified occurrences.

The proof of Theorem~\ref{thm:gz-f.correspondence}
relies on the notions of supported models and splitting sets
that we lift from standard ASP to programs with conditional constraints atoms, as stated below.
For clarity,
we abuse notation and
let~$\Headn(r)$ and $\Body(r)$ stand for formulas~$\bigvee \Headn(r)$ and~$\bigwedge \Body(r)$, respectively.
%
\begin{definition}[Supported models]
\label{def:supported}
A variable $x \in \X$ is \emph{supported} wrt a program~$\Pi$ and a valuation~$v$,
if there is a rule~$r \in \Pi$ and a constraint atom~$c\in\Headp(r)$ satisfying the following conditions:
\begin{enumerate}
\item $x \in \varsp{c}$,
\label{def:supported:cond:1}
\item $v \not\modelss c'$ for every $c' \in\Headp(r)$ such that $x \notin \varsp{c'}$,
\label{def:supported:cond:2}
\item $v \modelss\Body(r)$ and $v\not\modelss\Headn(r)$.
\label{def:supported:cond:3}
\end{enumerate}
A model~$v$ of a program~$\Pi$ is \emph{supported}
if every variable that is not undefined is supported wrt.~$\Pi$ and~$v$.
\end{definition}
\begin{proposition}\label{prop:supported}
Any stable model of a program is supported.
\end{proposition}
\begin{definition}[Splitting]\label{def:splitting.set}
  A set of variables \mbox{$U \subseteq \mathcal{X}$} is a \emph{splitting set} of a program $\Pi$,
  if for any rule $r$ in $\Pi$ one of the following conditions holds:
  \begin{enumerate}
  \item $\vars{r} \subseteq U$,
    \label{item:1:def:splitting}
  \item $\varsp{\Headp(r)} \cap U = \emptyset$
    \label{item:3:def:splitting}%
  \end{enumerate}
  We define a \emph{splitting} of
  $\Pi$ as a pair $\tuple{B_U(\Pi),T_U(\Pi)}$
  satisfying $B_U(\Pi) \cap T_U(\Pi) = \emptyset$,
  $B_U(\Pi) \cup T_U(\Pi) = \Pi$,
  all rules in $B_U(\Pi)$ satisfy~\ref{item:1:def:splitting}.\ and all rules in $T_U(\Pi)$ satisfy~\ref{item:3:def:splitting}.
\end{definition}
%
Given a program~$\Pi$, a splitting set~$U$ of~$\Pi$ and a valuation~$v$,
we denote by $E_U(\Pi,v)$ the program obtained
by replacing each variable $x \in U$ in $T_U(\Pi)$ by $v(x)$.
We denote by $\overline{U} \eqdef \X \setminus U$ the complement of~$U$.
\begin{proposition}{\label{thm:splitting}}
  Given program~$\Pi$ with splitting set ${U \subseteq \mathcal{X}}$,
  a valuation~$v$ is a stable model of~$\Pi$
  iff
  $\restr{v}{U}$ is a stable model of $B_U(\Pi)$ and
  $\restr{v}{\; \overline{U}}$ is a stable model of $ E_U(\Pi,\restr{v}{U})$.
\end{proposition}
%

\section{A generalization of Ferraris' semantics}

In this section,
we show that, when restricted to~\mbox{$\f$-semantics},
our approach amounts to a conservative extension of the reduct-based semantics introduced by~\citet{ferraris11a}.
Under that approach, 
a classical interpretation is a stable model of a formula
if it is a subset minimal classical model of the reduct wrt that interpretation.
The reduct of a formula wrt an interpretation is obtained by replacing all maximum subformulas not classically satisfied by the interpretation by $\bot$.
We now adapt those notions to \HTC.

Given a denotation $\den{\cdot}$,
a valuation $t$ \emph{classically satisfies} a formula $\varphi$,
written $t \modelscl \varphi$,
if the following conditions hold:
\begin{enumerate}
\item $t \not\modelscl\bot$
\item $t \modelscl c \text{ if } t\in \den{\evalcl{t}(c)}$
\item $t \modelscl \varphi \land \psi \text{ if }  t \modelscl \varphi \text{ and }  t \modelscl \psi$
\item $t \modelscl \varphi \lor \psi \text{ if }  t \modelscl \varphi \text{ or }  t \modelscl \psi$
\item $t \modelscl \varphi \rightarrow \psi
\text{ if }t \not\modelscl \varphi \text{ or }t \modelscl \psi$
\end{enumerate}
%
We say that a valuation~$v$ is a \emph{classical model} of theory~$\Gamma$ when $v \modelscl \varphi$ for all formulas~${\varphi \in \Gamma}$.

\begin{observation}
\label{lem:htcc:modelscl}
For any interpretation $\langle t,t \rangle $, formula $\varphi$
and \kmapping~$\sem$,
we have $\langle t,t \rangle \modelss \varphi$ iff $t\modelscl \varphi$
\end{observation}
\begin{definition}[Reduct]
The \emph{reduct} of a formula $\varphi$ wrt an interpretation $t$, written $\varphi^t$, is defined as the expression:
\begin{itemize}
\item $\bot$ if $t\not\modelscl \varphi$ for any formula~$\varphi$,

\item $c[ s_1^t,s_2^t,\dots]$
if ${t \modelscl c[ s_1,s_2,\dots]}$ for any constraint atom~${c \in \C}$ where ${s_1,s_2,\dots}$ are all conditional terms in~$c$ and $s_i^t \eqdef {\cterm{s}{s'}{\varphi^t}}$ for each $s_i={\cterm{s}{s'}{\varphi}}$.

\item $\varphi_1^t \otimes \varphi_2^t$
if $t \modelscl \varphi_1 \otimes \varphi_2$ with $\otimes \in \{\land,\lor,\rightarrow\}$
\end{itemize}
\end{definition}

The reduct of a theory $\Gamma$ is defined as~${\Gamma^t \eqdef \{ \varphi^t \mid \varphi \in \Gamma\}}$.
A valuation~$v$ is called a \mbox{$\ff$-stable} model of~$\Gamma$ iff it is a $\subseteq$-minimal model of~$\Gamma^t$.

An aggregate atom of the form of
\begin{gather}\
\op\langle \ (s_1 | 0_\op \!:\! \varphi_1), \ (s_2 | 0_\op \!:\! \varphi_2), \dotsc \rangle \prec s_0
  \label{eq:agg.asp}
\end{gather}
can be seen as a generalization of aggregate atoms as defined in~\cite{ferraris11a}
in three ways.
First, it permits applying the operation~$\op$ to both finite or infinite collections
of elements.
Second, it allows operations~$\op$ that may or may not depend on the order of the elements in the collection.
And third, and more important for our purposes, it allows each~$s_i$ to be any expression involving constraint variables rather than just numbers.
%
%
The following result shows that
the application of our reduct to an aggregate of the form of~\eqref{eq:agg.asp}
produces a straightforward generalization of \citeauthor{ferraris11a}' reduct.
%
\begin{proposition}
\label{prop:htcc:aggreduct}
Given an aggregate~$A$ of the form~\eqref{eq:agg.asp}
and a valuation~$t$,
it follows that
\begin{gather*}
A^t \!=\! \begin{cases}
\bot &\text{if } t \not\models c
\\
\op\langle (s_1 | 0_\op \!:\! \varphi_1^t), \, (s_2|0_\op \!:\! \varphi_2^t), \dotsc \rangle \prec s_0
&\text{otherwise}
\end{cases}
\end{gather*}
\end{proposition}

Let us now enunciate the main result of this section.
%
\begin{theorem}
\label{thm:htcc:eqstab}
A valuation is a $\f$-stable model of a theory~$\Gamma$ iff it is an $\ff$-stable model of~$\Gamma$.
\end{theorem}
%

\section{Discussion}\label{sec:discussion}

\HTC\ is a logic for capturing non-monotonic constraint theories that permits assigning default values to constraint variables.
Since ASP is a special case of this logic,
it provides a uniform framework for integrating ASP and CP on the same semantic footing.
We elaborate on this logic by incorporating \emph{constraint terms}.
A notion that allows us to treat linear constraints, conditional expressions and aggregates in a uniform way.
In particular,
this allows us to introduce assignments for aggregate expressions.
We also present a new semantics for conditional expressions in which their result is always defined~($\f$)
and show that, when combined with an appropriate definition of aggregates,
it leads to a generalization of the semantics by~\citet{ferraris11a}.
Recall that this semantics is the foundation for aggregates in the system~\clingo.

Interestingly,
for programs stratified on aggregates,
we can translate aggregates using the $\f$~principle into conditional constraints under the \emph{vicious circle principle}.
%
Then, we can leverage our previous results and translate these constructs into the language of CASP solvers.
%
As a reminder, the fragment covered by the ASP~Core~2 semantics~\cite{aspcore2} only allows for stratified aggregates.

Ongoing work is directed towards an implementation of a hybrid variant of~\clingo\ based on the framework developed here.
For solving programs with non-stratified aggregates,
we are looking into extending the notions of unfounded-sets~\cite{gerosc91a} and loop formulas~\cite{linzha04a} to programs with constraint variables.


\bibliographystyle{aaai}

\appendix
\clearpage
\section{Proofs of results}
\label{sec:proofs}
\begin{proof}{Proposition~\ref{prop:htcc:gzprime}}
Let $\modelsgz'$ be the satisfaction relation resulting from replacing Condition~\ref{item:htc:atom} with Condition~\ref{item:htc:atom}'.
In the following proof, we focus on Condition~\ref{item:htc:atom}' as the induction base.
The other cases are identical to $\modelsgz$. The full proof is obtained via structural induction.
\begin{itemize}

\item  $\langle h,t\rangle \modelsgz c$ iff $\langle h,t\rangle \modelsgz' c$ for $c\in\mathcal{C}$:
     \begin{align}
                   &\langle h,t\rangle\modelsgz c\label{prop:htcc:gzprime:eq1}\\
     \text{iff}\,  &h\in \den{\eval{h}{t}(c)}\text{ and }t\in \den{\eval{t}{t}(c)}\label{prop:htcc:gzprime:eq2}\\
     \text{iff}\,  &h\in \den{\eval{h}{t}(c)}\label{prop:htcc:gzprime:eq3}\\
     \text{iff}\,  &\langle h,t\rangle\modelsgz' c \label{prop:htcc:gzprime:eq4}
     \end{align}
    Equivalence between \eqref{prop:htcc:gzprime:eq1} and \eqref{prop:htcc:gzprime:eq2} holds by definition of the satisfaction relation.
    Equivalence between \eqref{prop:htcc:gzprime:eq2} and \eqref{prop:htcc:gzprime:eq3} holds by definition of the evaluation function and conditions \ref{den:prt:2} and \ref{den:prt:3} on denotations.
    More specifically, $h\in \den{\eval{h}{t}(c)}$ implies $t\in \den{\eval{t}{t}(c)}$ since for any conditional term $s=\cterm{s'}{s''}{\varphi}$ in $c$, either
	    \begin{enumerate}
	     \item $\tuple{h,t}\models\varphi$, then by persistence since $\varphi$ is condition-free $\tuple{t,t}\models\varphi$ and therefore $\evalgz{h}{t}(\ctermm{s'}{s''}{\varphi})=\evalgz{t}{t}(\ctermm{s'}{s''}{\varphi})=s'$,

	     \item $\tuple{t,t}\not\models\varphi$, and therefore
       $\evalgz{h}{t}(\ctermm{s'}{s''}{\varphi})=\evalgz{t}{t}(\ctermm{s'}{s''}{\varphi})=s''$,

	     \item or $\evalgz{h}{t}(\ctermm{s'}{s''}{\varphi})=\undefined$.
	    \end{enumerate}
	 For (a) and (b) evaluation of $s$ is identical.
	 For (c), we have $c[s/\undefined]$ when evaluating in $\tuple{h,t}$ 
	 and either $c[s/s']$ or $c[s/s'']$ when evaluating in $\tuple{t,t}$. 
	 In both cases, $h\in\den{c[s/\eval{h}{t}(s)]}$ implies $h\in\den{c[s/\eval{t}{t}(s)]}$
	 due to Condition~\ref{den:prt:3} on denotations. 
	 Thus, we have $h\in\den{\eval{h}{t}(c)}$ implies $h\in\den{\eval{t}{t}(c)}$,
	 and ultimately $t\in\den{\eval{t}{t}(c)}$ due to Condition~\ref{den:prt:2} on denotations.

   Implication between \eqref{prop:htcc:gzprime:eq3} and \eqref{prop:htcc:gzprime:eq4} holds by definition of the satisfaction relation.
   
\end{itemize}
\end{proof}
\begin{proof}{Proposition~\ref{prop:htcc:evalt}}
   We have
   \begin{align}
               &\tuple{h,t} \modelss c\label{prop:htcc:evalt:eq:1}\\
    \text{iff }&h\in\den{\evals{\sem}{h}{t}(c)} \text{ and } t\in\den{\evals{\sem}{t}{t}(c)}\label{prop:htcc:evalt:eq:2}\\
    \text{iff }&h\in \den{\evals{\sem}{h}{t}(c)} \text{ and }t\in \den{\evals{\sem}{h}{t}(c)}\nonumber\\  
    &\text{ and }\tuple{t,t} \models  \evalcl{t}(c)\label{prop:htcc:evalt:eq:3}\\
    \text{iff }&\tuple{h,t} \models \evals{\sem}{h}{t}(c) \text{ and } \tuple{h,t} \models\neg\neg  \evalcl{t}(c)\label{prop:htcc:evalt:eq:4}\\
    \text{iff }&\tuple{h,t} \models \evals{\sem}{h}{t}(c) \wedge \neg\neg  \evalcl{t}(c)\label{prop:htcc:evalt:eq:5}
   \end{align}

   Equivalence between \eqref{prop:htcc:evalt:eq:1} and \eqref{prop:htcc:evalt:eq:2} holds by definition of the satisfaction relation.
   Equivalence between \eqref{prop:htcc:evalt:eq:2} and \eqref{prop:htcc:evalt:eq:3} holds since 
   $h\in \den{\evals{\sem}{h}{t}(c)}$ implies $t\in\den{\evals{\sem}{h}{t}(c)}$ by Condition~\ref{den:prt:2} on denotations,
   $\evals{\sem}{t}{t}(c)=\evalcl{t}(c)$ and $t\in\den{\evalcl{t}(c)}$ iff $\tuple{t,t} \models  \evalcl{t}(c)$ 
   by definition of the evaluation function for total models and satisfaction relation.
   Note that $\evals{\sem}{h}{t}(c)$ and $\evalcl{t}(c)$ are condition-free constraint atoms and as such we can use $\models$.
   Equivalence between \eqref{prop:htcc:evalt:eq:3} and \eqref{prop:htcc:evalt:eq:4} holds by definition of the satisfaction relation,
   and since implication from \eqref{prop:htcc:evalt:eq:3} to \eqref{prop:htcc:evalt:eq:4} holds as 
   $\tuple{t,t}\models \evalcl{t}(c)$ implies $\tuple{h,t}\not\models \neg \evalcl{t}(c)$ by contraposition of Proposition~\ref{prop:htcc:properties}.\ref{item:2:prop:htcc:properties}
   and also $\tuple{t,t}\not\models \neg \evalcl{t}(c)$ by definition of the satisfaction relation for implication,
   which in turn implies $\tuple{h,t} \models\neg\neg  \evalcl{t}(c)$,
   and implication from \eqref{prop:htcc:evalt:eq:4} to \eqref{prop:htcc:evalt:eq:3} holds 
   as $\tuple{h,t}\models\neg\neg  \evalcl{t}(c)$ implies $\tuple{t,t}\not\models \neg \evalcl{t}(c)$ and ultimately $\tuple{t,t}\models \evalcl{t}(c)$
   again by definition of the satisfaction relation for implication.
   Finally, equivalence between \eqref{prop:htcc:evalt:eq:4} and \eqref{prop:htcc:evalt:eq:5} hold by definition of the satisfaction relation for conjunction.
\end{proof}
\begin{proof}{Proposition~\ref{prop:htcc:persistence}}
In the following, we focus on Condition~\ref{item:htc:atom} of the satisfaction relation,
as the proposition was proven in \cite{cakaossc16a} for $HT_c$ without conditional terms and the satisfaction is identical but for Condition~\ref{item:htc:atom}.
The full proof is obtained via structural induction with the remaining cases.
     \begin{align}
     \text{Assume } &\langle h,t\rangle\modelss c\label{prop:htcc:persistence:eq1}\\
     \Rightarrow\, &h\in \den{\evals{\sem}{h}{t}(c)} \text{ and } t\in \den{\evals{\sem}{t}{t}(c)}\label{prop:htcc:persistence:eq2}\\
     \Rightarrow\, &t \in \den{\evals{\sem}{t}{t}(c)}\label{prop:htcc:persistence:eq3}\\
     \Rightarrow\, &\langle t,t\rangle\modelss c \label{prop:htcc:persistence:eq4}
     \end{align}
    Implications between \eqref{prop:htcc:persistence:eq1} and \eqref{prop:htcc:persistence:eq2} 
    and between \eqref{prop:htcc:persistence:eq3} and \eqref{prop:htcc:persistence:eq4} hold by definition of the satisfaction relation.
\end{proof}

\begin{proof}{Proposition~\ref{prop:htcc:properties}.\ref{item:2:prop:htcc:properties}}
  We proof $\langle h,t\rangle \modelss \varphi\rightarrow\bot$ implies $\langle t,t\rangle \not\modelss \varphi$ for any formula $\varphi$.
     \begin{align*}
      \text{Assume }&\langle h,t\rangle\modelss\varphi\rightarrow \bot\\
      \Rightarrow&\langle t,t\rangle\modelss\varphi\rightarrow \bot\text{ due to Proposition~\ref{prop:htcc:properties}}\\
      \Rightarrow&\langle t,t\rangle\not\modelss\varphi\text{ or }\langle t,t\rangle\modelss\bot\\
      \Rightarrow& \langle t,t\rangle\not\modelss\varphi
     \end{align*}
  We proof $\langle t,t\rangle \not\modelss \varphi$ implies $\langle h,t\rangle \modelss \varphi\rightarrow\bot$ for $c\in\mathcal{C}$.
     \begin{align*}
      \text{Assume }&\langle t,t\rangle\not\modelss \varphi\\
      \Rightarrow&\langle h,t \rangle\not\modelss \varphi\text{ due to contraposition of Proposition~\ref{prop:htcc:properties}}\\
      \Rightarrow&\langle h,t\rangle\not\modelss \varphi\text{ or }\langle h,t\rangle\modelss\bot\text{ and }\langle t,t\rangle\not\modelss \varphi\text{ or }\langle t,t\rangle\modelss\bot\\
      \Rightarrow&\langle h,t \rangle\modelss \varphi\rightarrow\bot
     \end{align*}
\end{proof}

\begin{lemma}\label{lem:aux:prop:htcc:properties}
Let~$\varphi$ be any formula and~$\tuple{h,t}$ be an interpretation.
Let~$\tuple{H,T}$ be the \HT interpretation build as follows
\begin{align*}
T &= \{c\in \mathcal{C} \mid t\in\den{\evals{\sem}{t}{t}(c)}\}
\\
H &= \{c\in \mathcal{C} \mid h\in\den{\evals{\sem}{h}{t}(c)}\text{ and }t\in\den{\evals{\sem}{t}{t}(c)}\}
\end{align*}
Then,
$\tuple{h,t}\modelss\varphi$ iff  $\tuple{H,T}\models\varphi$.
\end{lemma}

\begin{proof}{Lemma~\ref{lem:aux:prop:htcc:properties}}
Then, $\tuple{H,T}$ is a valid \HT\ interpretation
due to $H\subseteq T$,
as every element in $H$ fulfills $t\in\den{\evals{\sem}{t}{t}(c)}\}$ and is therefore also in $T$.
$T$ might have more elements as it drops the condition $h\in\den{\evals{\sem}{h}{t}(c)}$.

We proof property~\ref{item:3:prop:htcc:properties} in Proposition~\ref{prop:htcc:properties} by prooving $\tuple{h,t}\modelss\varphi$ iff  $\tuple{H,T}\models\varphi$ for any interpretation~$\tuple{h,t}$ and formula~$\varphi$, and thus tautologies are preserved between \HT\ and \HTC.
Since satisfaction relations are identical except for Condition~\ref{item:htc:atom}, we focus on $\varphi=c$ for $c\in\mathcal{C}$ as the induction base.
The other cases follow by induction since they are identical between~\HT\ and \HTC.

We proof $\tuple{h,t}\modelss c$ iff  $\tuple{H,T}\models c$ for $c\in\mathcal{C}$:
     \begin{align}
                   &\langle h,t\rangle\modelss c\label{prop:htcc:properties:eq5}\\
     \text{iff }\, &h\in \den{\evals{\sem}{h}{t}(c)}\text{ and }t\in \den{\evals{\sem}{t}{t}(c)}\label{prop:htcc:properties:eq6}\\
     \text{iff }\, &c \in H\label{prop:htcc:properties:eq7}\\
     \text{iff }\, &\tuple{H,T}\models c \label{prop:htcc:properties:eq8}
     \end{align}
Equivalence between \eqref{prop:htcc:properties:eq5} and \eqref{prop:htcc:properties:eq6} holds by definition of the satisfaction relation.
Equivalence between \eqref{prop:htcc:properties:eq6} and \eqref{prop:htcc:properties:eq7} holds by definition of $H$.
Equivalence between \eqref{prop:htcc:properties:eq7} and \eqref{prop:htcc:properties:eq8} holds by definition of the satisfaction relation.
\end{proof}

\begin{proof}{Proposition~\ref{prop:htcc:properties}.\ref{item:3:prop:htcc:properties}}
Let~$\varphi[\overline{a}]$ be a \HT~tautology
and let $\varphi[\overline{a}/\overline{c}]$ be the result of uniformly replacing the atoms
${\overline{a}=(a_1,\dots,a_n)}$ in $\varphi$ by a tuple of constraint atoms ${\overline{c} = (c_1,\dots,c_n)}$.
Take any interpretation~$\tuple{h,t}$ and let~$\tuple{H,T}$ be an \HT-interpretation build as in Lemma~\ref{lem:aux:prop:htcc:properties}.
Let $\tuple{H',T'}$ be an interpretation build as follows:
\begin{align*}
T' &= \{ a_i \mid c_i \in T \}
\\
H' &= \{ a_i \mid c_i \in H \}
\end{align*}
Then,
${\tuple{H',T'} \models \varphi[\overline{a}]}$
\\iff
${\tuple{H,T} \models \varphi[\overline{a}/\overline{c}]}$ (Signature independence)
\\iff
${\tuple{h,t} \models \varphi[\overline{a}/\overline{c}]}$ (Lemma~\ref{lem:aux:prop:htcc:properties}).
\\
Then,
since~$\varphi[\overline{a}]$ is a tautology,
we get that $\tuple{h,t} \models \varphi[\overline{a}/\overline{c}]$ for every~$\tuple{h,t}$ and, thus,~$\varphi[\overline{a}/\overline{c}]$ is a tautology.
Finally,
note that if~$\varphi[\overline{a}]$ be a \HT~tautology,
so is~$\varphi[\overline{a}/\overline{\beta}]$
where $\varphi[\overline{a}/\overline{\beta}]$ is the result of uniformly replacing the atoms
${\overline{a}=(a_1,\dots,a_n)}$ in $\varphi$ by a tuple of formulas ${\overline{\beta} = (\beta_1,\dots,\beta_n)}$
and, thus, the result holds for any \HTC formula~so is~$\varphi[\overline{a}/\overline{\alpha}]$.
\end{proof}

\begin{proof}{Proposition~\ref{prop:htcc:termequality}}
If ${s = s'}$ and ${s' = s''}$ and ${s = s''}$ are constraint atoms and they satisfy
${v \in \den{ s = s' }}$ and ${v \in \den{ s' = s'' }}$,
then there is ${d \in \D}$ such that ${v \in \den{ s = d }}$ and ${v \in \den{ s' = d }}$ and ${v \in \den{ s'' = d }}$
and, thus, 
${v \in \den{ s = s'' }}$
because ${(s = s'')[s''/d] = (s = d)}$.
\\[10pt]
Similarly,
if ${s = s'}$ and ${s' = s}$ are constraint atoms and we have
${v \in \den{ s = s' }}$,
then
there is ${d \in \D}$ such that ${v \in \den{ s = d }}$ and ${v \in \den{ s' = d }}$
and, thus, 
${v \in \den{ s' = s }}$
because ${(s' = s)[s/d] = (s' = d)}$.
\\[10pt]
Suppose, for the sake of contradiction, that~${v \in \den{ \undefined = s }}$.
Then, from~\eqref{den:prt:eq.4},
there is $d \in \D$ such that
${v \in \den{ \undefined = d}}$ and ${v \in \den{ s' = d}}$.
Furthermore, since~$\vars{\undefined = d} = \emptyset$,
from~\eqref{den:prt:1},
we can assume with out loss of generality that
${v(x) = \undefined}$.
From~\eqref{den:prt:3}
this implies that
${v \in \den{x = d}}$ holds because
${(x = d)[x/\undefined] = (\undefined = d)}$.
This is a contradiction with~\eqref{den:prt:eq.2}.
\end{proof}
\begin{proof}{Proposition~\ref{prop:htcc:eval.valuation}}
We have
\begin{align}
 &v \in \den{\evals{\sem}{h}{t}(c)}\label{prop:htcc:eval.valuation:eq:1}\\
 \text{iff }& v \in \den{c[\evals{\sem}{h}{t}(s_1)][\evals{\sem}{h}{t}(s_2)]\dots}\label{prop:htcc:eval.valuation:eq:2}\\
 \text{iff }& v \in \den{c[v_{\tuple{h,t}}^\sem(s_1)][v_{\tuple{h,t}}^\sem(s_2)]\dots}\label{prop:htcc:eval.valuation:eq:3}\\
 \text{iff }& v \in \den{v_{\tuple{h,t}}^\sem(c)}\label{prop:htcc:eval.valuation:eq:4}
\end{align}
where $s_i$ are all occurrences of terms in $c$.
Equivalence between \eqref{prop:htcc:eval.valuation:eq:1} and \eqref{prop:htcc:eval.valuation:eq:2} holds by definition of the evaluation function applied to a constraint atom. 
Equivalence between \eqref{prop:htcc:eval.valuation:eq:2} and \eqref{prop:htcc:eval.valuation:eq:3} holds by definition of evaluation function, $v_{\tuple{h,t}}$ 
and conditions \ref{den:prt:3}, \ref{den:prt:8} and \ref{den:prt:10} on denotations.
For any term $s_i$ in $c$, 
either $v \in\den{\evals{\sem}{h}{t}(s_i) = d}$ for some $d \in \D$,
then $v_{\tuple{h,t}}^\sem(s_1)=d$,
and therefore $v\in\den{c}$ iff $v\in\den{c[\evals{\sem}{h}{t}(s_i)/d]}$,
or $v \not\in \den{ \evals{\sem}{h}{t}(s_i) = d}$ for all $d \in \D$,
implying $v \not\in \den{ \evals{\sem}{h}{t}(s_i) = \evals{\sem}{h}{t}(s_i)}$,
and therefore $v\in\den{c}$ iff $v\in\den{c[\evals{\sem}{h}{t}(s_i)/\undefined]}$.
Finally, equivalence between \eqref{prop:htcc:eval.valuation:eq:3} and \eqref{prop:htcc:eval.valuation:eq:4} holds by definition of application of $v_{\tuple{h,t}}^\sem$ to constraint atoms. 

\end{proof}
\begin{proof}{Proposition~\ref{prop:term:persistence}}
If $h_{\tuple{h,t}}^\gz(s)\neq \undefined$ then $h\in\den{\evals{\gz}{h}{t}(s)=d}$ for some $d\in\mathcal{D}$ by definition of $v_{\tuple{h,t}}^\gz$.
For any conditional term $s'$ occuring in $s$,
we have either $\evals{\gz}{h}{t}(s')=\undefined\neq\evals{\f}{h}{t}(s')$ or $\evals{\gz}{h}{t}(s')=\evals{\f}{h}{t}(s')$ by definition of the evaluation functions.
Then $h\in\den{\evals{\gz}{h}{t}(s)=d}$ implies $h\in\den{\evals{\f}{h}{t}(s)=d}$ 
as in the former case $\undefined$ is replaced by some term and the implication holds by Condition~\ref{den:prt:3} on denotations,
or the evaluation is identical.
Furthermore, $h\in\den{\evals{\gz}{h}{t}(s)=d}$ implies $t\in\den{\evalcl{t}(s)=d}$ by Condition~\ref{den:prt:2} on denotations
and since evlauation functions coincide for total models.
As a result, we have $h_{\tuple{h,t}}^\gz(s)=h_{\tuple{h,t}}^\f(s)=t(s)$ by definition of valuations applied to terms.
\end{proof}

\begin{proof}{Theorem~\ref{thm:grounding.agg.assigments}}
By definition,~\eqref{eq:1:thm:grounding.agg.assigments}
is the formula
\begin{align*}
\varphi \ \to \
(\df{ \tau }  \to (\op\langle s_1, s_2, \dotsc \rangle = x )
\end{align*}
which is HT-equivalent to
\begin{align}
\df{ \tau } \wedge \varphi \ \to \ \tau = x
  \label{eq:3:thm:grounding.agg.assigments}
\end{align}
Let~$\tuple{h,t}$ be a model of~\eqref{eq:3:thm:grounding.agg.assigments}
and assume that
\begin{gather}
\tuple{h,t} \modelss \varphi \wedge\tau = d
  \label{eq:4:thm:grounding.agg.assigments}
\end{gather}
Then, for all~$v \in \{h,t\}$,
we get that one of the following conditions hold:
\begin{enumerate}
\item $\tuple{v,t} \not\modelss \df{\tau}$,
  \label{item:1:thm:grounding.agg.assigments}
\item $\tuple{v,t} \not\modelss \varphi$, or
  \label{item:2:thm:grounding.agg.assigments}
\item $\tuple{v,t} \modelss \tau = x$
  \label{item:3:thm:grounding.agg.assigments}
\end{enumerate}
Note that~\eqref{eq:4:thm:grounding.agg.assigments}
implies~$\tuple{h,t}\modelss \df{\tau}$ and~$\tuple{h,t}\modelss\varphi$
which is a contradiction with conditions~\ref{item:1:thm:grounding.agg.assigments}
and~\ref{item:2:thm:grounding.agg.assigments}, respectively.
Furthermore,
condition~\ref{item:3:thm:grounding.agg.assigments}
implies that ${\{h,t \} \subseteq \den{ \tau = x}}$.
Then, since~${\{h,t \} \subseteq \den{ \tau = d}}$,
we get~${\{h,t \} \subseteq \den{ x = d}}$
and, thus,
that~${\tuple{h,t} \modelss x = d}$.
This implies that~$\tuple{h,t}$ is a model of all rules of the form of~\eqref{eq:2:thm:grounding.agg.assigments}
and, thus of~$\Gamma$.
\\[10pt]
The other way around.
Let~$\tuple{h,t}$ be a model of~$\Gamma$
and assume
\begin{gather}
\tuple{h,t} \modelss \df{\tau} \wedge \varphi
  \label{eq:5:thm:grounding.agg.assigments}
\end{gather}
Since $\tuple{h,t}$ is a model of~$\Gamma$,
we get that, for every~${d \in \D}$ and every~${v \in \{h,t\}}$,
one of the following conditions hold:
\begin{enumerate}[ start=6]
\item $\tuple{v,t} \not\modelss \tau = d$,
  \label{item:6:thm:grounding.agg.assigments}
\item $\tuple{v,t} \not\modelss \varphi$, or
  \label{item:7:thm:grounding.agg.assigments}
\item $\tuple{v,t} \modelss x = d$
  \label{item:8:thm:grounding.agg.assigments}
\end{enumerate}
It is easy to see that~\eqref{item:6:thm:grounding.agg.assigments}
is a contradiction with the assumption~\eqref{eq:5:thm:grounding.agg.assigments}.
Furthermore, 
$\tuple{h,t} \modelss\df{\tau}$
implies that there is some~${d \in \D}$ such that
${v \in \den{\tau = d}}$ with~$v \in \{h,t\}$.
This is a contradiction with condition~\ref{item:6:thm:grounding.agg.assigments}.
Consequently,
we get that~${v \in \den{x = d}}$ with ${v \in \{v,t\}}$.
This plus
${v \in \den{\tau = d}}$
imply 
${v \in \den{\tau = x}}$
and, thus, that
$\tuple{h,t}$
is a model of~\eqref{eq:3:thm:grounding.agg.assigments} and~\eqref{eq:1:thm:grounding.agg.assigments}.
\end{proof}
\begin{proof}{Proposition~\ref{prop:htcc:aggtermeval}}
Assume $v_{\tuple{h,t}}^\sem(A)\neq\undefined$ for aggregate term $A=\op\langle s_1, s_2, \dotsc \rangle$.
Then we have
\begin{align}
  v_{\tuple{h,t}}^\sem(A)=&d\label{eq:htcc:aggtermeval:1}\\
  =&v(\evals{\sem}{h}{t}(A))\label{eq:htcc:aggtermeval:2}\\
  =&v(\op\langle \evals{\sem}{h}{t}(s_1), \evals{\sem}{h}{t}(s_2), \dotsc \rangle)\label{eq:htcc:aggtermeval:3}\\
  =&\hat{\op}(\theta)\label{eq:htcc:aggtermeval:4}\\
  =&\hat{\op}(\theta_{\tuple{h,t}}^\sem)\label{eq:htcc:aggtermeval:5}
\end{align}
where $d\in\mathcal{D}$, 
$\theta:\mathbb{N}^+\to\mathcal{D}$ with $\theta(i)=v(s_i')=v(\evals{\sem}{h}{t}(s_i))$ for $i\geq 1$,
and $\theta_{\tuple{h,t}}^\sem: \mathbb{N}^+ \to \D$ with $\theta_{\tuple{h,t}}^\sem(i)=v_{\tuple{h,t}}^\sem(s_i)$ for $i\geq 1$.
Equivalence between \eqref{eq:htcc:aggtermeval:1} and \eqref{eq:htcc:aggtermeval:2} holds by definition of $v_{\tuple{h,t}}^\sem(A)$,
and since $v_{\tuple{h,t}}^\sem(A)=d\neq\undefined$,
$v\in\den{\evals{\sem}{h}{t}(A)=d}$ and by definition of denotation for equality,
$v(\evals{\sem}{h}{t}(A))=d\neq\undefined$.
Equivalence between \eqref{eq:htcc:aggtermeval:2} and \eqref{eq:htcc:aggtermeval:3} holds by definition of application of the evaluation function to terms.
Equivalence between \eqref{eq:htcc:aggtermeval:3} and \eqref{eq:htcc:aggtermeval:4} holds by definition of valuations applied to condition-free aggregate terms,
since $\op\langle \evals{\sem}{h}{t}(s_1), \evals{\sem}{h}{t}(s_2), \dotsc \rangle)$ is condition-free.
And finally, equivalence between \eqref{eq:htcc:aggtermeval:4} and \eqref{eq:htcc:aggtermeval:5}
holds since $\theta=\theta_{\tuple{h,t}}^\sem$ as $v_{\tuple{h,t}}^\sem(s_i)=v(\evals{\sem}{h}{t}(s_i))\neq\undefined$ by definition of $v_{\tuple{h,t}}^\sem(s_i)$,
and the fact that no $v(\evals{\sem}{h}{t}(s_i))$ may be undefined because then $v(\op\langle \evals{\sem}{h}{t}(s_1), \evals{\sem}{h}{t}(s_2), \dotsc \rangle)=\undefined$ by definition of application of valuations to aggregate terms, 
which cannot be the case as $v(\op\langle \evals{\sem}{h}{t}(s_1), \evals{\sem}{h}{t}(s_2), \dotsc \rangle)=d\neq\undefined$.

Assume $v_{\tuple{h,t}}^\sem(A)=\undefined$,
then by definition there exists no $d\in\mathcal{D}$ such that $v\in\den{\evals{\sem}{h}{t}(A)=d}$,
which implies $v(\evals{\sem}{h}{t}(A))=\undefined$ as it is unequal to all domain elements.
As shown above $v(\evals{\sem}{h}{t}(A))=v(\op\langle \evals{\sem}{h}{t}(s_1), \evals{\sem}{h}{t}(s_2), \dotsc \rangle)$,
and $v(\op\langle \evals{\sem}{h}{t}(s_1), \evals{\sem}{h}{t}(s_2), \dotsc \rangle)=\undefined$ iff there exists a $v_{\tuple{h,t}}^\sem(s_i)=\undefined$ for $i\geq 1$ by definition of application of valuations to aggregate terms.

\end{proof} 

\begin{proof}{Proposition~\ref{prop:htcc:lintermeval}}
Assume $v_{\tuple{h,t}}^\sem(\alpha)\neq\undefined$ for linear term $\alpha=s_1+s_2+\dotsc$.
Then we have
\begin{align}
  v_{\tuple{h,t}}^\sem(\alpha)=&d\label{eq:htcc:lintermeval:1}\\
  =&v(\evals{\sem}{h}{t}(\alpha))\label{eq:htcc:lintermeval:2}\\
  =&v(\evals{\sem}{h}{t}(s_1)+\evals{\sem}{h}{t}(s_2)+ \dotsc)\label{eq:htcc:lintermeval:3}\\
  =&\sum_{i \geq 1} v_{\tuple{h,t}}^\sem(s_i)\label{eq:htcc:lintermeval:4}
\end{align}
where $d\in\mathcal{D}$, 
and $s_i$ are conditional linear terms occuring in $\alpha$.
Equivalence between \eqref{eq:htcc:lintermeval:1} and \eqref{eq:htcc:lintermeval:2} holds by definition of $v_{\tuple{h,t}}^\sem(\alpha)$,
and since $v_{\tuple{h,t}}^\sem(\alpha)=d\neq\undefined$,
$v\in\den{\evals{\sem}{h}{t}(\alpha)=d}$ and by definition of denotation for equality,
$v(\evals{\sem}{h}{t}(\alpha))=d\neq\undefined$.
Equivalence between \eqref{eq:htcc:lintermeval:2} and \eqref{eq:htcc:lintermeval:3} holds by definition of application of the evaluation function to linear terms.
Equivalence between \eqref{eq:htcc:lintermeval:3} and \eqref{eq:htcc:lintermeval:4} holds by definition of valuations applied to condition-free linear terms,
since $\evals{\sem}{h}{t}(s_1)+\evals{\sem}{h}{t}(s_2)+ \dotsc$ is condition-free.
Furthermore, $\evals{\sem}{h}{t}(s_i)\neq\undefined$ since $v(\evals{\sem}{h}{t}(s_1)+\evals{\sem}{h}{t}(s_2)+ \dotsc)\neq\undefined$
by definition of valuations applied to condition-free linear terms.
Therefore, $v(\evals{\sem}{h}{t}(s_i))=v_{\tuple{h,t}}^\sem(s_i)$ since $v_{\tuple{h,t}}^\sem(s_i)=d'$,
for which holds $v\in\den{\evals{\sem}{h}{t}(s_i)=d'}$ by definition of $v_{\tuple{h,t}}^\sem(s_i)$.

Assume $v_{\tuple{h,t}}^\sem(\alpha)=\undefined$,
then by definition there exists no $d\in\mathcal{D}$ such that $v\in\den{\evals{\sem}{h}{t}(\alpha)=d}$,
which implies $v(\evals{\sem}{h}{t}(\alpha))=\undefined$ as it is unequal to all domain elements.
As shown above $v(\evals{\sem}{h}{t}(\alpha))=v(\evals{\sem}{h}{t}(s_1)+\evals{\sem}{h}{t}(s_2)+ \dotsc)$,
and $v(\evals{\sem}{h}{t}(s_1)+\evals{\sem}{h}{t}(s_2)+ \dotsc)=\undefined$ if there exists a 
$v(\evals{\sem}{h}{t}(s_i))\not\in\mathbb{Z}$ by definition of application of valuations to condition-free linear terms.

\end{proof} 

\begin{proof}{Proposition~\ref{prop:htcc:linear.decomposition}.\ref{prop:htcc:linear.decomposition:1}}
We have 
\begin{align}
            &\tuple{h,t}\modelss\alpha=\beta\label{prop:htcc:linear.decomposition:1:eq:1}\\
 \text{iff }& h_{\tuple{h,t}}^\sem(\alpha)=h_{\tuple{h,t}}^\sem(\beta) \text{ and } t_{\tuple{t,t}}^\sem(\alpha)=t_{\tuple{t,t}}^\sem(\beta)\label{prop:htcc:linear.decomposition:1:eq:2}\\
 \text{iff }& h_{\tuple{h,t}}^\sem(\alpha)\leq h_{\tuple{h,t}}^\sem(\beta) \text{ and } h_{\tuple{h,t}}^\sem(\alpha)\geq h_{\tuple{h,t}}^\sem(\beta)\nonumber\\
            &\text{ and } t_{\tuple{t,t}}^\sem(\alpha)\leq t_{\tuple{t,t}}^\sem(\beta) \text{ and } t_{\tuple{t,t}}^\sem(\alpha)\geq t_{\tuple{t,t}}^\sem(\beta)\label{prop:htcc:linear.decomposition:1:eq:3}\\
 \text{iff }&\tuple{h,t}\modelss\alpha\leq\beta\text{ and }\tuple{h,t}\modelss\alpha\geq\beta\label{prop:htcc:linear.decomposition:1:eq:4}\\
 \text{iff }&\tuple{h,t}\modelss\alpha\leq\beta\land\alpha\geq\beta\label{prop:htcc:linear.decomposition:1:eq:5}   
\end{align}
Equivalence between \eqref{prop:htcc:linear.decomposition:1:eq:1} and \eqref{prop:htcc:linear.decomposition:1:eq:2} holds by Corollary~\ref{cor:htcc:lintermeval}.
Equivalence between \eqref{prop:htcc:linear.decomposition:1:eq:2} and \eqref{prop:htcc:linear.decomposition:1:eq:3} holds 
since $d=d'$ iff $d\leq d'$ and $d\geq d'$ for $d,d'\in\mathbb{Z}$,
and $s\in \mathbb{Z}$ for $s\in\{h_{\tuple{h,t}}^\sem(\alpha),t_{\tuple{t,t}}^\sem(\alpha),h_{\tuple{h,t}}^\sem(\beta),t_{\tuple{t,t}}^\sem(\beta)\}$
because otherwise $\tuple{h,t}\not\modelss\alpha=\beta$, $\tuple{h,t}\not\modelss\alpha\leq\beta$ and $\tuple{h,t}\not\modelss\alpha\geq\beta$ by definition of denotation for linear constraints.
Equivalence between \eqref{prop:htcc:linear.decomposition:1:eq:3} and \eqref{prop:htcc:linear.decomposition:1:eq:4} holds by Corollary~\ref{cor:htcc:lintermeval}.
Finally, equivalence between \eqref{prop:htcc:linear.decomposition:1:eq:4} and \eqref{prop:htcc:linear.decomposition:1:eq:5}
         holds by definition of the satisfaction relation for conjunction. 
\end{proof} 

\begin{proof}{Proposition~\ref{prop:htcc:linear.decomposition}.\ref{prop:htcc:linear.decomposition:2}}
We have 
\begin{align}
            &\tuple{h,t}\modelss\alpha<\beta\label{prop:htcc:linear.decomposition:2:eq:1}\\
 \text{iff }& h_{\tuple{h,t}}^\sem(\alpha)<h_{\tuple{h,t}}^\sem(\beta) \text{ and } t_{\tuple{t,t}}^\sem(\alpha)<t_{\tuple{t,t}}^\sem(\beta)\label{prop:htcc:linear.decomposition:2:eq:2}\\
 \text{iff }& h_{\tuple{h,t}}^\sem(\alpha)\leq h_{\tuple{h,t}}^\sem(\beta) \text{ and } h_{\tuple{h,t}}^\sem(\alpha)\neq h_{\tuple{h,t}}^\sem(\beta)\nonumber\\
            &\text{ and } t_{\tuple{t,t}}^\sem(\alpha)\leq t_{\tuple{t,t}}^\sem(\beta) \text{ and } t_{\tuple{t,t}}^\sem(\alpha)\neq t_{\tuple{t,t}}^\sem(\beta)\label{prop:htcc:linear.decomposition:2:eq:3}\\
 \text{iff }&\tuple{h,t}\modelss\alpha\leq\beta\text{ and }\tuple{h,t}\modelss\alpha\neq\beta\label{prop:htcc:linear.decomposition:2:eq:4}\\
 \text{iff }&\tuple{h,t}\modelss\alpha\leq\beta\land\alpha\neq\beta\label{prop:htcc:linear.decomposition:2:eq:5}   
\end{align}
Equivalence between \eqref{prop:htcc:linear.decomposition:2:eq:1} and \eqref{prop:htcc:linear.decomposition:2:eq:2} holds by Corollary~\ref{cor:htcc:lintermeval}.
Equivalence between \eqref{prop:htcc:linear.decomposition:2:eq:2} and \eqref{prop:htcc:linear.decomposition:2:eq:3} holds 
since $d<d'$ iff $d\leq d'$ and $d\neq d'$ for $d,d'\in\mathbb{Z}$,
and $s\in \mathbb{Z}$ for $s\in\{h_{\tuple{h,t}}^\sem(\alpha),t_{\tuple{t,t}}^\sem(\alpha),h_{\tuple{h,t}}^\sem(\beta),t_{\tuple{t,t}}^\sem(\beta)\}$
because otherwise $\tuple{h,t}\not\modelss\alpha<\beta$, $\tuple{h,t}\not\modelss\alpha\leq\beta$ and $\tuple{h,t}\not\modelss\alpha\neq\beta$ by definition of denotation for linear constraints.
Equivalence between \eqref{prop:htcc:linear.decomposition:2:eq:3} and \eqref{prop:htcc:linear.decomposition:2:eq:4} holds by Corollary~\ref{cor:htcc:lintermeval}.
Finally, equivalence between \eqref{prop:htcc:linear.decomposition:2:eq:4} and \eqref{prop:htcc:linear.decomposition:2:eq:5}
         holds by definition of the satisfaction relation for conjunction. 
\end{proof} 

\begin{proof}{Proposition~\ref{prop:htcc:linear.decomposition}.\ref{prop:htcc:linear.decomposition:3}}
We have 
\begin{align}
            &\tuple{h,t}\modelsgz\alpha<\beta\label{prop:htcc:linear.decomposition:3:eq:1}\\
 \text{iff }& h_{\tuple{h,t}}^\gz(\alpha)<h_{\tuple{h,t}}^\gz(\beta) \text{ and } t_{\tuple{t,t}}^\gz(\alpha)<t_{\tuple{t,t}}^\gz(\beta)\label{prop:htcc:linear.decomposition:3:eq:2}\\
 \text{iff }& h_{\tuple{h,t}}^\gz(\alpha)\leq h_{\tuple{h,t}}^\gz(\beta) \text{ and not } h_{\tuple{h,t}}^\gz(\alpha)\geq h_{\tuple{h,t}}^\gz(\beta)\nonumber\\
            &\text{ and } t_{\tuple{t,t}}^\gz(\alpha)\leq t_{\tuple{t,t}}^\gz(\beta) \text{ and not } t_{\tuple{t,t}}^\gz(\alpha)\geq t_{\tuple{t,t}}^\gz(\beta)\label{prop:htcc:linear.decomposition:3:eq:3}\\
 \text{iff }&\tuple{h,t}\modelsgz\alpha\leq\beta\text{ and }\tuple{h,t}\not\modelsgz\alpha\geq\beta\text{ and }\tuple{t,t}\not\modelsgz\alpha\geq\beta\label{prop:htcc:linear.decomposition:3:eq:4}\\
 \text{iff }&\tuple{h,t}\modelsgz\alpha\leq\beta\text{ and }\tuple{h,t}\modelsgz\alpha\geq\beta\rightarrow\bot\label{prop:htcc:linear.decomposition:3:eq:5}\\
 \text{iff }&\tuple{h,t}\modelsgz\alpha\leq\beta\land\neg(\alpha\geq\beta)\label{prop:htcc:linear.decomposition:3:eq:6}   
\end{align}
Equivalence between \eqref{prop:htcc:linear.decomposition:3:eq:1} and \eqref{prop:htcc:linear.decomposition:3:eq:2} holds by Corollary~\ref{cor:htcc:lintermeval}.
Equivalence between \eqref{prop:htcc:linear.decomposition:3:eq:2} and \eqref{prop:htcc:linear.decomposition:3:eq:3} holds 
since $d<d'$ iff $d\leq d'$ and not $d\geq d'$ for $d,d'\in\mathbb{Z}$,
and $s\in \mathbb{Z}$ for $s\in\{h_{\tuple{h,t}}^\gz(\alpha),t_{\tuple{t,t}}^\gz(\alpha),h_{\tuple{h,t}}^\gz(\beta),t_{\tuple{t,t}}^\gz(\beta)\}$
because otherwise $\tuple{h,t}\not\modelsgz\alpha<\beta$ and $\tuple{h,t}\not\modelsgz\alpha\geq\beta$ by definition of denotation for linear constraints.
Equivalence between \eqref{prop:htcc:linear.decomposition:3:eq:3} and \eqref{prop:htcc:linear.decomposition:3:eq:4} holds 
by definition of the satisfaction relation, Corollary~\ref{cor:htcc:lintermeval} and Proposition~\ref{prop:term:persistence}.
More specifically, $\tuple{t,t}\not\modelsgz\alpha\geq\beta$ implies that $t_{\tuple{t,t}}^\gz(\alpha)\geq t_{\tuple{t,t}}^\gz(\beta)$ does not hold,
which in turn implies that $h_{\tuple{h,t}}^\gz(\alpha)\geq h_{\tuple{h,t}}^\gz(\beta)$ does not hold,
because as mentioned before $h_{\tuple{h,t}}^\gz(\alpha)\neq\undefined$ and $h_{\tuple{h,t}}^\gz(\beta)\neq\undefined$
thus it follows that $h_{\tuple{h,t}}^\gz(\alpha)=t_{\tuple{t,t}}^\gz(\alpha)$ and $h_{\tuple{h,t}}^\gz(\beta)=t_{\tuple{t,t}}^\gz(\beta)$.
Finally, equivalences between \eqref{prop:htcc:linear.decomposition:3:eq:4} and \eqref{prop:htcc:linear.decomposition:3:eq:5}
         and \eqref{prop:htcc:linear.decomposition:3:eq:5} and \eqref{prop:htcc:linear.decomposition:3:eq:6}
         hold by definition of the satisfaction relation for implication and conjunction, respectively. 
\end{proof} 

\begin{proof}{Proposition~\ref{prop:htcc:linear.decomposition}.\ref{prop:htcc:linear.decomposition:4}}
We have 
\begin{align}
            &\tuple{h,t}\modelsgz\alpha\neq\beta\label{prop:htcc:linear.decomposition:4:eq:1}\\
 \text{iff }& h_{\tuple{h,t}}^\gz(\alpha)\neq h_{\tuple{h,t}}^\gz(\beta) \text{ and } t_{\tuple{t,t}}^\gz(\alpha)\neq t_{\tuple{t,t}}^\gz(\beta)\label{prop:htcc:linear.decomposition:4:eq:2}\\
 \text{iff }& h_{\tuple{h,t}}^\gz(\alpha)< h_{\tuple{h,t}}^\gz(\beta) \text{ or } h_{\tuple{h,t}}^\gz(\alpha)> h_{\tuple{h,t}}^\gz(\beta),\nonumber\\
            &\text{ and } t_{\tuple{t,t}}^\gz(\alpha)< t_{\tuple{t,t}}^\gz(\beta) \text{ or } t_{\tuple{t,t}}^\gz(\alpha)> t_{\tuple{t,t}}^\gz(\beta)\label{prop:htcc:linear.decomposition:4:eq:3}\\
 \text{iff }& h_{\tuple{h,t}}^\gz(\alpha)< h_{\tuple{h,t}}^\gz(\beta) \text{ and } t_{\tuple{t,t}}^\gz(\alpha)< t_{\tuple{t,t}}^\gz(\beta)\nonumber\\
            &\text{ or } h_{\tuple{h,t}}^\gz(\alpha)> h_{\tuple{h,t}}^\gz(\beta) \text{ and } t_{\tuple{t,t}}^\gz(\alpha)> t_{\tuple{t,t}}^\gz(\beta)\label{prop:htcc:linear.decomposition:4:eq:4}\\
 \text{iff }&\tuple{h,t}\modelsgz\alpha<\beta\text{ or }\tuple{h,t}\modelsgz\alpha>\beta\label{prop:htcc:linear.decomposition:4:eq:5}\\
 \text{iff }&\tuple{h,t}\modelsgz\alpha<\beta\lor\alpha>\beta\label{prop:htcc:linear.decomposition:4:eq:6}   
\end{align}
Equivalence between \eqref{prop:htcc:linear.decomposition:4:eq:1} and \eqref{prop:htcc:linear.decomposition:4:eq:2} holds by Corollary~\ref{cor:htcc:lintermeval}.
Equivalence between \eqref{prop:htcc:linear.decomposition:4:eq:2} and \eqref{prop:htcc:linear.decomposition:4:eq:3} holds 
since $d\neq d'$ iff $d<d'$ or $d>d'$ for $d,d'\in\mathbb{Z}$,
and $s\in \mathbb{Z}$ for $s\in\{h_{\tuple{h,t}}^\gz(\alpha),t_{\tuple{t,t}}^\gz(\alpha),h_{\tuple{h,t}}^\gz(\beta),t_{\tuple{t,t}}^\gz(\beta)\}$
because otherwise $\tuple{h,t}\not\modelsgz\alpha<\beta$, $\tuple{h,t}\not\modelsgz\alpha<\beta$ and $\tuple{h,t}\not\modelsgz\alpha\neq\beta$ by definition of denotation for linear constraints.
Equivalence between \eqref{prop:htcc:linear.decomposition:4:eq:3} and \eqref{prop:htcc:linear.decomposition:4:eq:4} holds 
by Proposition~\ref{prop:term:persistence}.
More specifically, because as mentioned before $h_{\tuple{h,t}}^\gz(\alpha)\neq\undefined$ and $h_{\tuple{h,t}}^\gz(\beta)\neq\undefined$,
it follows that $h_{\tuple{h,t}}^\gz(\alpha)=t_{\tuple{t,t}}^\gz(\alpha)$ and $h_{\tuple{h,t}}^\gz(\beta)=t_{\tuple{t,t}}^\gz(\beta)$.
Therefore, if $h_{\tuple{h,t}}^\gz(\alpha)< h_{\tuple{h,t}}^\gz(\beta)$ and $t_{\tuple{t,t}}^\gz(\alpha)< h_{\tuple{t,t}}^\gz(\beta)$ hold,
$h_{\tuple{h,t}}^\gz(\alpha)>h_{\tuple{h,t}}^\gz(\beta)$ and $t_{\tuple{t,t}}^\gz(\alpha)> h_{\tuple{t,t}}^\gz(\beta)$ are false and vice-versa.
Equivalence between \eqref{prop:htcc:linear.decomposition:4:eq:4} and \eqref{prop:htcc:linear.decomposition:4:eq:5} holds by Corollary~\ref{cor:htcc:lintermeval}.
Finally, equivalence between \eqref{prop:htcc:linear.decomposition:4:eq:5} and \eqref{prop:htcc:linear.decomposition:4:eq:6}
holds by definition of the satisfaction relation for disjunction. 
\end{proof} 

\begin{proof}{Theorem~\ref{thm:htcc:linconstagg}}
 We proof the theorem by showing that for any interpretation $\tuple{h,t}$,
 we have $\tuple{h,t}\modelsp{\sem} \mathit{sum}\agg{ \tau_1, \tau_2, \dotsc}\prec s$ iff $\tuple{h,t}\modelsp{\pi(\sem)} \pi(\mathit{sum}\agg{ \tau_1, \tau_2, \dotsc})\prec s$.
 We do this by proving that elements of the respective sums are identical and therefore the resulting values.
 By definition of evaluation function and denotation for condition-free aggregates, 
 $h\in\den{\evals{\sem}{h}{t}(\htag{sum}{ \tau_1, \tau_2, \dotsc } \prec s)}$ 
 iff $\widehat{sum}(\theta) \prec h(s)$,
 where $\theta:\mathbb{N}^+\to\mathcal{D}$ maps $i\in \mathbb{N}^+$ to  $h(\evals{\sem}{h}{t}(\tau_i))$
   for an interpretation $\tuple{h,t}$.
 Similarly, by definition of $\pi$, evaluation function and denotation for condition-free aggregates and valuations applied to linear terms, 
 $h\in\den{\evals{\pi(\sem)}{h}{t}(\pi(\htag{sum}{ \tau_1, \tau_2, \dotsc}) \prec s))}$ 
 iff 
 \begin{align*}
 h\in\den{\evals{\pi(\sem)}{h}{t}(\pi(\tau_1) + \pi(\tau_2), \dotsc \prec s)}
 \end{align*}  
 iff
 \begin{align*}
 h(\evals{\pi(\sem)}{h}{t}(\pi(\tau_1)) +h(\evals{\pi(\sem)}{h}{t}(\pi(\tau_2))+\dots \prec h(s)
 \end{align*}  
   for an interpretation $\tuple{h,t}$.
   First of all, note that no $h(\evals{\sem}{h}{t}(\tau_i))=\undefined$ or $h(\evals{\pi(\sem)}{h}{t}(\pi(\tau_i))=\undefined$,
   because then $\tuple{h,t}\not\modelsp{\sem} \mathit{sum}\agg{ \tau_1, \tau_2, \dotsc}\prec s$
   or $\tuple{h,t}\not\modelsp{\pi(\sem)} \pi(\mathit{sum}\agg{ \tau_1, \tau_2, \dotsc})\prec s$, respectively,
   by definition of evaluation for aggregate terms and linear terms and denotation for aggregates and linear constraints. 
  As relation and $s$ are identical, 
  we only have to show 
 \begin{align*}
  \widehat{sum}(&\theta)= h(\evals{\pi(\sem)}{h}{t}(\pi(\tau_1)) +h(\evals{\pi(\sem)}{h}{t}(\pi(\tau_2))+\dots
 \end{align*}  
 By definition of $\widehat{sum}$, we have
 \[
  \widehat{sum}(\theta)=\sum \{ \, h(\evals{\sem}{h}{t}(\tau_i))  \mid i \in\mathbb{N}^+ \text{ and } h(\evals{\sem}{h}{t}(\tau_i))\in\mathbb{Z} \, \}.
 \]
 For the translation, we have
 \begin{align*}
  &h(\evals{\pi(\sem)}{h}{t}(\pi(\tau_1)) +h(\evals{\pi(\sem)}{h}{t}(\pi(\tau_2))+\dots\\
  =&\sum \{ \, h(\evals{\pi(\sem)}{h}{t}(\pi(\tau_i)))  \mid i \in\mathbb{N}^+ \text{ and } h(\evals{\sem}{h}{t}(\tau_i))\in\mathbb{Z} \, \}
 \end{align*}  
 because each $\pi(\tau_i)$ is conditioned either by $\isint{\tau_i}$, if $\tau_i$ is a finite basic linear term,
 or $\isint{s_i}$ if $\tau_i$ is of the form $s_i:\varphi$,
 and therefore $h(\evals{\sem}{h}{t}(\tau_i))\not\in\mathbb{Z}$ implies $h(\evals{\pi(\sem)}{h}{t}(\pi(\tau_i)))=0$ and we can safely remove those elements from the sum.
 As such, we only have to show $h(\evals{\sem}{h}{t}(\tau_i))=h(\evals{\pi(\sem)}{h}{t}(\pi(\tau_i)))$
 for $h(\evals{\sem}{h}{t}(\tau_i))\in\mathbb{Z}$.
 For finite basic linear terms, this follows immediately as $h(\evals{\sem}{h}{t}(\tau_i))=h(\tau_i)=h(\evals{\pi(\sem)}{h}{t}(\tau_i:\isint{\tau_i}))$
 since $h(\tau_i)\in\mathbb{Z}$.
 For conditional terms $s_i:\varphi$, we have $h(\evals{\sem}{h}{t}(\cterm{s_i}{0}{\varphi\land\df{s_i}}))\in\mathbb{Z}$
 if either $\tuple{h,t}\not\models \varphi$,
 then $h(\evals{\sem}{h}{t}(\cterm{s_i}{0}{\varphi\land\df{s_i}}))=0$
 and $h(\evals{\pi(\sem)}{h}{t}(\cterm{s_i}{0}{\varphi\land\isint{s_i}}))=0$,
 or $\tuple{h,t}\models \varphi$,
 then $h(\evals{\sem}{h}{t}(\cterm{s_i}{0}{\varphi\land\df{s_i}}))=h(s_i)$
 and $h(\evals{\pi(\sem)}{h}{t}(\cterm{s_i}{0}{\varphi\land\isint{s_i}}))=h(s_i)$,
 because $h(s_i)\in\mathbb{Z}$ and $\sem$ and $\pi(\sem)$ is identical for $\cterm{s_i}{0}{\varphi\land\df{s_i}}$ and $\cterm{s_i}{0}{\varphi\land\isint{s_i}}$
 by definition of $\pi$.
  
 As such, we have
 \begin{align*}
 &\tuple{h,t}\modelss \mathit{sum}\agg{ \tau_1, \tau_2, \dotsc}\prec s\\
\text{iff }&h\in\den{\evals{\sem}{h}{t}(\mathit{sum}\agg{ \tau_1, \tau_2, \dotsc}\prec s)}\\
           &\text{ and } t\in\den{\evals{\sem}{t}{t}(\mathit{sum}\agg{ \tau_1, \tau_2, \dotsc}\prec s)}\\
\text{iff }&h\in\den{\evals{\pi(\sem)}{h}{t}(\pi(\mathit{sum}\agg{ \tau_1, \tau_2, \dotsc})\prec s)}\\
           &\text{ and } t\in\den{\evals{\pi(\sem)}{t}{t}(\pi(\mathit{sum}\agg{ \tau_1, \tau_2, \dotsc})\prec s)}\\
\text{iff }&\tuple{h,t}\modelsp{\pi(\sem)} \pi(\mathit{sum}\agg{ \tau_1, \tau_2, \dotsc})\prec s)
 \end{align*}
\end{proof} 

\begin{lemma}\label{lem:supported.aux}
Let $\tuple{t,t}$ be an equilibrium model of some logic program~$\Pi$ without assignment rules,
$x \in \X$ be some variable which is defined in~$t$
and
$\tuple{h,t}$ be some interpretation with
$h(x) = \undefined$
and
$h(y) = t(y)$ for every variable~$y \in \X \setminus\{x\}$.
Then,
there is a rule~$r \in \Pi$ 
and a constraint atom~$c\in\Headp(r)$ satisfying the following conditions:
\begin{enumerate}
\item $x \in \var(c)$,
 \label{lem:supported:cond:1}
\item $\tuple{t,t} \not\modelss c'$ for every $c' \in\Headp(r)$ such that $x \notin \var(c')$,
 \label{lem:supported:cond:2}
\item $\tuple{h,t} \modelss \Body(r)$ and $\tuple{t,t} \not\modelss \Headn(r)$
 \label{lem:supported:cond:3}
\end{enumerate}
\end{lemma}

\begin{proof}{Lemma~\ref{lem:supported.aux}}
 We show that $\tuple{h,t}\modelss\Pi$ whenever one of the conditions \ref{lem:supported:cond:1}-\ref{lem:supported:cond:3} is not fulfilled for $x$,
 thus a contradiction follows with $\tuple{t,t}$ being an equilibrium model,
 and a rule $r$ fulfilling all conditions has to exist.
 \begin{itemize}
   \item[\ref{lem:supported:cond:1}] Assume there exists no $c$ with $x\in\vars{c}$ for $c\in\Headp(r)$ and \mbox{$r\in\Pi$}, 
   then $\tuple{t,t} \modelss \Head(r)$ implies $\tuple{h,t} \modelss \Head(r)$ for every rule~\mbox{$r \in \Pi$}.
   This is obvious in case that $x\not\in\vars{\Head(r)}$ as it has no impact on satisfaction due to Condition~\ref{den:prt:1} on denotations.
   If $x\in \vars{\neg c'}$ for some $\neg c'\in \Headn(r)$, $\tuple{t,t}\modelss c'\to \bot$ implies $\tuple{h,t}\modelss c'\to \bot$
   since $\tuple{t,t}\not\modelss c'$ implies $\tuple{h,t}\not\modelss c'$ by contraposition of Proposition~\ref{prop:htcc:persistence}.
   Furthermore, since $\tuple{t,t}$ is a model of~$\Pi$, it follows that
   $\tuple{t,t} \modelss \Body(r) \to \Head(r)$ and, thus,
   either
   $\tuple{t,t} \modelss \Head(r)$
   or
   $\tuple{t,t} \not\modelss\Body(r)$.
   As we have seen above, the former implies that
   $\tuple{h,t} \modelss \Head(r)$.
   From Proposition~\ref{prop:htcc:persistence},
   the latter implies
   $\tuple{h,t} \not\modelss\Body(r)$.
   Hence
   $\tuple{h,t} \modelss \Body(r) \to \Head(r)$.
   This implies that $\tuple{h,t}\modelss \Pi$ which contradicts the fact that $\tuple{t,t}$ is an equilibrium model of $\Pi$ and 
   therefore a rule has to exists with $x$ in the positive head.
   
   \item[\ref{lem:supported:cond:2}] Since we proved Lemma~\ref{lem:supported.aux}.\ref{lem:supported:cond:1},
   we only need to consider rules $r$ with $x\in\vars{\Headp(r)}$.
   
   Assume there exists a $c'\in \Headp(r)$ with $x\not\in\vars{c'}$ such that $\tuple{t,t}\modelss c'$ for all rules $r\in\Pi$ with $x\in\vars{\Headp(r)}$, 
   then we know $\tuple{t,t}\modelss \Head(r)$ and $\tuple{h,t}\modelss \Head(r)$,
   due to definition of satisfaction relation for disjunction and Condition~\ref{den:prt:1} for the denotation.
   Thus rule $r$ is fulfilled regardless of the body due to definition of satisfaction relation for implication.
   It follows that $\tuple{h,t}\modelss\Pi$ which again contradicts $\tuple{t,t}$ being an equilibrium model and therefore there is at least a rule with $x$ in the positive head
   where no other positive head atom not containing $x$ is satisfied.
   \item[\ref{lem:supported:cond:3}] 
   We only have to consider rules $r\in\Pi$ with $x\in\vars{\Headp(r)}$ and there exists no $c'\in \Headp(r)$ with $x\not\in\vars{c'}$ and $\tuple{t,t}\modelss c'$
   as shown above.
   Assume $\tuple{h,t}\not\modelss\Body(r)$,
   then rule $r$ is satisfied regardless of the head.
   Similarly, assume $\tuple{t,t}\modelss\Headn(r)$.
   This implies $\tuple{h,t}\modelss\Headn(r)$,
   as there exists a literal $\neg c'\in \Headn(r)$ for which holds $\tuple{t,t}\modelss c' \to \bot$ by definition of disjunction,
   in turn implying $\tuple{h,t}\modelss c' \to \bot$ since $\tuple{t,t}\not\modelss c'$ implies $\tuple{h,t}\not\modelss c'$ by contraposition of Proposition~\ref{prop:htcc:persistence}.
   Therefore, $\tuple{h,t}\modelss\Head(r)$ and rule $r$ is satisfied regardless of the body.
   In both cases, $\tuple{h,t}\modelss r$ and in turn $\tuple{h,t}\modelss\Pi$,
   our final contradiction to $\tuple{t,t}$ being an equilibrium model.
 \end{itemize}
\end{proof}

\begin{proposition}\label{prop:supported:woassignments}
Every stable model of a logic program without assignment rules is also supported.
\end{proposition}

\begin{proof}{Proposition~\ref{prop:supported:woassignments}}
 Proposition~\ref{prop:supported:woassignments} follows directly from Lemma~\ref{lem:supported.aux}, 
 as conditions \ref{def:supported:cond:1} and \ref{def:supported:cond:2} in the definition of supported are identical to conditions \ref{lem:supported:cond:1} and \ref{lem:supported:cond:2} in Lemma~\ref{lem:supported.aux},
 and Condition~\ref{lem:supported:cond:3} implies $\tuple{t,t}\modelss \Body(r)$ by Proposition~\ref{prop:htcc:persistence},
 as well as $\tuple{t,t}\not\modelss \Headn(r)$.
 Thus, every stable model $t$ is supported.
\end{proof} 

\begin{proof}{Proposition~\ref{prop:supported}}
First note that,
from Theorem~\ref{thm:grounding.agg.assigments},
we can construct a logic program $\Pi'$ without assignment rules
such that $\Pi\equiv_\sem\Pi'$ for every \kmapping~$\sem$
by replacing rules of the form ${x:=s\leftarrow \varphi}$ in $\Pi$ 
with rules of the form ${x=d\leftarrow \varphi\land s=d}$ in $\Pi'$.
Then,
Proposition~\ref{prop:supported:woassignments},
implies that there is a rule in~$\Pi'$ satisfying conditions of Definition~\ref{def:supported}.
If such a rule also belongs to~$\Pi$, the result follows immediately.
Otherwise,
such a rule is of the form~${x=d\leftarrow \varphi\land s=d}$
and corresponds to a rule of the form~${x:=s\leftarrow \varphi}$ in~$\Pi$.
Conditions~\ref{def:supported:cond:1} and~\ref{def:supported:cond:2} are immediately satisfied.
For condition~\ref{def:supported:cond:1},
we have ${v \modelss \varphi \wedge s=d}$
which implies
${v \modelss B(r)}$
because ${B(r) = \varphi \wedge \df{s}}$
and
${v \modelss (s=d)}$ implies ${v \modelss \df{s}}$.
\end{proof} 

  For splitting, we need the following notation.
  Given  an interpretation $\tuple{h,t}$,
  by $\restr{\tuple{h,t}}{U}$
  we deonte the interpretation
  $\tuple{\restr{h}{U}, \restr{t}{U}}$.

  \begin{observation}{\label{prop:splitting.aux}}
  Let ${U \subseteq \mathcal{X}}$ be some set of atoms and let ${\Pi = \Pi_1 \cup \Pi_2}$ be some without assignment rules such that
  $\mathit{var}(\Pi_1) \subseteq U$ and $\mathit{var}(\Pi_2) \subseteq \overline{U}$.
  Then, any interpretation $\tuple{h,t}$ is an (equilibrium) model of $\Pi$ iff both
  $\restr{\tuple{h,t}}{U}$ is an (equilibrium) model of $\Pi_1$ and
  $\restr{\tuple{h,t}}{\overline{U}}$ is an (equilibrium) model of~$\Pi_2$.
  \end{observation}

  \begin{observation}\label{lem:splitting.aux20}
  Let $\varphi$ be some theory, $\sem$ be some \kmapping, $\tuple{h,t}$ be some total interpretation and $x \in \X$ be some variable such that $h(x) = t(x)$.
  Let $\varphi'$ be the result of replacing some occurrence of $x$ by $t(x)$.
  Then, $\tuple{h,t} \modelss \varphi$ iff $\tuple{h,t} \modelss \varphi'$.
  \end{observation}

  \begin{lemma}{\label{lem:splitting.aux21}}
  Let $\Pi$ be a without assignment rules with splitting set~$U \subseteq \mathcal{X}$.
  Let $\tuple{h,t}$ be a model of $B_U(\Pi)$
  such that $\tuple{t,t}$ is a model of $T_U(\Pi)$.
  Let $h'$ be a valuation such that
  ${\restr{h'\!}{U} = \restr{h}{U}}$
  and ${\restr{h'\!}{\overline{U}} = \restr{t}{\overline{U}}}$.
  Then,
  $\tuple{h',t}$ is a model of~$\Pi$.
  \end{lemma}

  \begin{proof}{Lemma~\ref{lem:splitting.aux21}}
  Since every ${r\in B_U(\Pi)}$ satisfies ${\vars{r} \subseteq U}$
  and ${\restr{h'\!}{U} = \restr{h}{U}}$,
  we immediately get that
  $\tuple{h',t}$ satisfies all rules in~$B_U(\Pi)$.
  \\[3pt]
  Pick now any rule~${r \in T_U}$.
  Given that~$\tuple{t,t}$ is a model of~$T_U(\Pi)$,
  we get that either
  ${\tuple{t,t} \modelss \Headp(r)}$
  or
  ${\tuple{t,t} \modelss \Headn(r)}$
  or
  ${\tuple{t,t} \not\modelss \Body(r)}$
  hold.
  We reason by cases:
  \begin{itemize}
  \item since
  ${\vars{\Headp(r)} \subseteq \overline{U}}$
  and
  ${\restr{h'\!}{\overline{U}} = \restr{t}{\overline{U}}}$,
  we get that
  ${\tuple{t,t} \modelss \Headp(r)}$
  implies
  ${\tuple{h',t} \modelss \Head(r)}$.
  
  \item Since $\Headn(r)$ is a negative formula,
  we get that ${\tuple{t,t} \modelss \Headn(r)}$
  implies
  ${\tuple{h',t} \modelss \Headn(r)}$
  because 
  
  \item From Proposition~\ref{prop:htcc:persistence},
  we get that
  ${\tuple{t,t} \not\modelss \Body(r)}$
  implies
  ${\tuple{h',t} \not\modelss \Body(r)}$
  \end{itemize}
  In all three cases we get that ${\tuple{h',t} \modelss r}$
  and, therefore, we get that ${\tuple{h',t}}$
  satisfies all rules in~$T_U(\Pi)$
  and, consequently, in~$\Pi$.
  \end{proof}

  \begin{lemma}{\label{lem:splitting.aux22}}
  Let $\Pi$ be a without assignment rules with splitting set~${U \subseteq \mathcal{X}}$,
  $\tuple{t,t}$ be a total interpretation.
  Let ${r \in T_U(\Pi)}$ be a rule, ${x \in U}$ be some variable,
  and let $r'$ be the result of of replacing some occurrence of~$x$ by $t(x)$.
  Then, $\tuple{t,t}$ is an equilibrium model of~$\Pi$
  iff $\tuple{t,t}$ is an equilibrium model of~$(\Pi \setminus \{ r \}) \cup \{ r' \}$.
  \end{lemma}

  \begin{proof}{Lemma~\ref{lem:splitting.aux22}}
  Let $\Pi' = (\Pi \setminus \{ r \}) \cup \{ r' \}$.
  From Observation~\ref{lem:splitting.aux20},
  we get that $\tuple{t,t}$ is a model of~$\Pi$
  iff $\tuple{t,t}$ is a model of~$\Pi'$.
  \\[5pt]
  Assume that~$\tuple{t,t}$ is an equilibrium model of~$\Pi$ (resp.~$\Pi'$).
  Then, to show that $\tuple{t,t}$ is an equilibrium model of~$\Pi'$ (resp.~$\Pi$)
  we need to prove that no interpretation~$\tuple{h,t}$ with ${h \subset t}$
  is a model of~$\Pi'$ (resp.~$\Pi$).
  \\[5pt]
  Let~$\tuple{h,t}$ be any such interpretation
  and suppose, for the sake of contradiction, that it is a model of~$\Pi'$ (resp.~$\Pi$).
  Then~$\tuple{h,t}$ is also a model of~$\Pi \setminus\{r\}$
  and, from Lemma~\ref{lem:splitting.aux21},
  we get that there is a model~$\tuple{h',t}$ of~$\Pi$ (resp.~$\Pi'$)
  such that~$\restr{h}{U} = \restr{h'\!}{U}$.
  Since~$\tuple{t,t}$ is an equilibrium model of~$\Pi$ (resp.~$\Pi'$),
  it follows that $h' = t$
  and, thus, we get that~$\restr{h}{U} = \restr{t}{U}$.
  Since ${x \in U}$,
  this implies ${h(x) = t(x)}$
  and, from Observation~\ref{lem:splitting.aux20},
  we get that
  ${\tuple{h,t} \models r}$
  iff
  ${\tuple{h,t} \models r'}$.
  Hence,
  $\tuple{h,t}$ is a model of~$\Pi$ (resp.~$\Pi'$)
  which is a contradiction with the assumption that~$\tuple{t,t}$ is an equilibrium model of~$\Pi$ (resp.~$\Pi'$).
  Consequently,
  $\tuple{h,t}$ cannot be model of~$\Pi'$ (resp.~$\Pi$).
  As a result,~$\tuple{t,t}$ is an equilibrium model of~$\Pi$ (resp.~$\Pi'$).
  \end{proof}

  \begin{proposition}{\label{prop:splitting.aux2}}
  Let $\Pi$ be a without assignment rules with splitting set~${U \subseteq \mathcal{X}}$.
  Then, any total interpretation~$\tuple{t,t}$ is an equilibrium model of~$\Pi$
  iff $\tuple{t,t}$ is an equilibrium model of~$B_U(\Pi) \cup E_U(\Pi,t)$.
  \end{proposition}

  \begin{proof}{Proposition~\ref{prop:splitting.aux2}}
  Follows directly by induction using Lemma~\ref{lem:splitting.aux22}.
  \end{proof}

  \begin{proof}{Theorem~\ref{thm:splitting}}
  Assume first that $\tuple{t,t}$ is an equilibrium model of~$\Pi$.
  From Proposition~\ref{prop:splitting.aux2},
  it follows that that $\tuple{t,t}$ is an equilibrium model of~$B_U(\Pi) \cup E_U(\Pi,t)$.
  Furthermore, by construction, we can see that
  $\mathit{var}(B_U(\Pi)) \subseteq U$ and $\mathit{var}(E_U(\Pi,t)) \subseteq \overline{U}$.
  Then, from Observation~\ref{prop:splitting.aux},
  it follows that that
  $\restr{\tuple{t,t}}{U}$ is an equilibrium model of $B_U(\Pi)$ and
  $\restr{\tuple{t,t}}{\overline{U}}$ is an equilibrium model of $E_U(\Pi,t)$.
  The other way around is symetric.
  \end{proof}

  \begin{theorem}{\label{thm:gz-f.correspondence:woassignments}}
  Let $\Pi$ be some program without assignment rules which is stratified on
  some occurrence of a conditional term~$s$
  and $\sem$ and~$\sem'$ be two evaluation mappings that agree on all occurrences of conditional terms but~$s$.
  Then, the $\sem$-stable models and the $\sem'$-stable models of~$\Pi$ coincide.
  \end{theorem}

  \begin{proof}{Theorem~\ref{thm:gz-f.correspondence:woassignments}}
  Let $U$ be
  $$\{ x \in \mathcal{X} \mid \lambda(x) \leq \lambda(y) \text{ for some } y \in \vars{\varphi}\ \}$$
  Let~$r \in \Pi$ be any rule.
  If there is $x \in (\Headp(r) \cap U)$,
  then we get that $\lambda(y) \leq \lambda(x)$ for every variable~$y \in \vars{r}$
  and, therefore, $\vars{r} \subseteq U$.
  Consequently, $U$ is a splitting set of~$\Pi$

  From Theorem~\ref{thm:splitting},
  this implies that
  any total interpretation~$\tuple{t,t}$ is an equilibrium model of~$\Pi$
  iff 
  $\restr{\tuple{t,t}}{U}$ is an equilibrium model of $B_U(\Pi)$ and
  $\restr{\tuple{t,t}}{\overline{U}}$ is an equilibrium model of $E_U(\Pi,\restr{t}{U})$.

  Let $r \in \Pi$  (resp.~$r\in \Pi'$) be rule in which the conditional expression~$\tau$ occurs.
  Then,
  $\lambda(x) > \lambda(y)$ for all variables ${x \in \vars{\Headp(r)}}$
  and~${y \in \vars{\varphi}}$.
  This implies that~$\Headp(r) \cap U = \emptyset$ and, thus, that $r \in T_U(\Pi)$ (resp.~$r \in T_U(\Pi')$).
  Hence, we immediately get that
  the equilibrium models of~$B_U(\Pi)$ and~$B_U(\Pi')$
  coincide.

  Let $r' =  E_U(r,\restr{t}{U})$ and~$\tau'$ be the result of applying this transformation to~$\tau$.
  Then, since~$\vars{\varphi} \subseteq U$,
  we get that every variable in~$\varphi$ has been replaced by its value in~$r'$.
  This implies~$\evals{\sem}{h}{t}{\tau'}$ and~$\evals{\sem'}{h}{t}{\tau'}$
  coincide
  and, as a result,
  the equilibrium models of~$E_U(\Pi,\restr{t}{U})$ and~$E_U(\Pi',\restr{t}{U})$
  also coincide.
  Consequently, $\Pi$ and~$\Pi'$ have the same equilibrium models.
  \end{proof}
  
  \begin{proof}{Theorem~\ref{thm:gz-f.correspondence}}
  First note that,
  from Theorem~\ref{thm:grounding.agg.assigments},
  we can construct a logic program $\Pi'$ without assignment rules
  such that $\Pi\equiv_\sem\Pi'$ for every \kmapping~$\sem$
  by replacing rules of the form ${x:=s\leftarrow \varphi}$ in $\Pi$ 
  with rules of the form ${x=d\leftarrow \varphi\land s=d}$ in $\Pi'$.
  It is easy to see that, if~$\Pi$ is stratified in some occurrence, then so it is~$\Pi'$.
  Hence, the result follows directly from Theorem~\ref{thm:gz-f.correspondence:woassignments}.
  \end{proof}

\begin{proof}{Theorem~\ref{thm:gz-f.correspondence2}}
First note that, if a program is tight, then it is stratified on all occurrences not in the scope of negation.
Then, let~$\sem$ and~$\sem'$ be two \kmappings agreeing on all occurrences in the scope of negation.
From Theorem~\ref{thm:gz-f.correspondence},
we immediately get that the $\sem$ and~$\sem'$-stable models coincide.
Furthermore, from Proposition~\ref{prop:htcc:properties} and Observation~\ref{obs:eval.total},
we have~${\tuple{h,t} \modelss \neg\varphi}$ iff ${\tuple{t,t} \modelss \neg\varphi}$
iff ${\tuple{t,t} \models_{\sem''} \neg\varphi}$ for any \kmapping~$\sem''$.
Then the result follows.
\end{proof}
\begin{proof}{Proposition~\ref{prop:htcc:aggreduct}}
Note that the otherwise case is just an abbreviation for
\begin{gather*}\label{eq:cfree.aggreagate}
\op\langle \cterm{s_1}{0_\op}{\varphi_1^t}, \cterm{s_2}{0_\op}{\varphi_2^t}, \dotsc \rangle \prec s_0
\end{gather*}
and its easy to see that this is the result of applying the reduct to~\eqref{eq:agg.asp.expanded}
when this is satisfied by~$t$.
\end{proof}

\begin{lemma}\label{lem:htcc:confree}
 For any interpretation $\langle h,t \rangle$ and condition-free formula $\varphi$,
 we have $\langle h,t \rangle\modelsf\varphi$ iff $h\modelscl\varphi^t$
\end{lemma}

\begin{proof}{Lemma~\ref{lem:htcc:confree}}
    We proof Lemma~\ref{lem:htcc:confree} via induction over the structure of the formula.
    \paragraph{Induction base} For condition-free constraint atom $c$, we get
      \begin{align}
                & \langle h,t\rangle\modelsf c \label{eq:lem:61} \\
\text{iff } & h\in\den{\evalf{h}{t}(c)}\text{ and }t\in\den{\evalf{t}{t}(c)}\label{eq:confreeht} \\
\text{iff } & h\in\den{\evalcl{h}(c)}\text{ and }t\in\den{\evalcl{t}(c)}\label{eq:confreefer} \\
\text{iff } & h\modelscl c\text{ and }t\modelscl c \label{eq:lem:62} \\
\text{iff } & h\modelscl c^t  \label{eq:lem:63} 
      \end{align} 
      Equivalence between \eqref{eq:lem:61} and \eqref{eq:confreeht} holds by definition of the satisfaction relation.
      Equivalence between \eqref{eq:confreeht} and \eqref{eq:confreefer} holds by definition of the evaluation function. 
          since $\evalf{v}{v'}(c)=\evalcl{v''}(c)=c$ for any condition-free constraint $c$ and valuations $v,v',v''$.
      Equivalence between \eqref{eq:confreefer} and \eqref{eq:lem:62} holds by definition of the classical satisfaction relation.
      Equivalence between \eqref{eq:lem:62} and \eqref{eq:lem:63} holds by definition of the reduct 
      since $c^t=c$ for any condition-free constraint atom $c$ such that $t\modelscl c$,
      which in turn has to hold since otherwise $h$ is no classical model of $c^t$.
        
    \paragraph{Induction step} For formula $\varphi_1 \land \varphi_2$, we get
          \begin{align}
               &\langle h,t \rangle \modelsf \varphi_1 \land \varphi_2\label{eq:lem:71}\\
\text{iff }&\langle h,t \rangle \modelsf \varphi_1\text{ and }\langle h,t \rangle \modelsf \varphi_2\label{eq:lem:72}\\
\text{iff }&h\modelscl \varphi_1^t\text{ and }h \modelscl \varphi_2^t\label{eq:lem:73}\\
\text{iff }&h\modelscl (\varphi_1\land \varphi_2)^t\label{eq:lem:74}
          \end{align}
     Equivalence between \eqref{eq:lem:71} and \eqref{eq:lem:72} holds by definition of the satisfaction relation.
     Equivalence between \eqref{eq:lem:72} and \eqref{eq:lem:73} holds by induction hypothesis.
     Equivalence between \eqref{eq:lem:73} and \eqref{eq:lem:74} holds by definitions of the classical satisfaction relation and reduct.
     \paragraph{Induction step} For formula $\varphi_1 \lor \varphi_2$, we get
          \begin{align}
               &\langle h,t \rangle \modelsf \varphi_1 \lor \varphi_2\label{eq:lem:81}\\
\text{iff }&\langle h,t \rangle \modelsf \varphi_1\text{ or }\langle h,t \rangle \modelsf \varphi_2\label{eq:lem:82}\\
\text{iff }&h\modelscl \varphi_1^t\text{ or }h \modelscl \varphi_2^t\label{eq:lem:83}\\
\text{iff }&h\modelscl (\varphi_1\lor \varphi_2)^t\label{eq:lem:84}
          \end{align}
     Equivalence between \eqref{eq:lem:81} and \eqref{eq:lem:82} holds by definition of the satisfaction relation.
     Equivalence between \eqref{eq:lem:82} and \eqref{eq:lem:83} holds by induction hypothesis.
     Equivalence between \eqref{eq:lem:83} and \eqref{eq:lem:84} holds by definitions of the classical satisfaction relation and reduct.
     \paragraph{Induction step} For formula $\varphi_1 \rightarrow \varphi_2$, we get
          \begin{align}
               &\langle h,t \rangle \modelsf \varphi_1 \rightarrow \varphi_2\label{eq:lem:91}\\
\text{iff }&\langle h,t \rangle \not\modelsf \varphi_1\text{ or }\langle h,t \rangle \modelsf \varphi_2\text{, and }\langle t,t \rangle \models \varphi_1 \rightarrow \varphi_2\label{eq:lem:92}\\
\text{iff }&h \not\modelscl \varphi_1^t\text{ or }h \modelscl \varphi_2^t\text{, and }t \modelscl \varphi_1 \rightarrow \varphi_2\label{eq:lem:93}\\
\text{iff }&h\modelscl (\varphi_1 \rightarrow \varphi_2)^t\label{eq:lem:94}
          \end{align}
      Equivalence between \eqref{eq:lem:91} and \eqref{eq:lem:92} holds by definition of the satisfaction relation and Proposition~\ref{prop:htcc:persistence}.
      Equivalence between \eqref{eq:lem:92} and \eqref{eq:lem:93} holds by induction hypothesis and Lemma~\ref{lem:htcc:modelscl}.
      Equivalence between \eqref{eq:lem:93} and \eqref{eq:lem:94} holds by definitions of the classical satisfaction relation and reduct.
\end{proof}
\begin{lemma}\label{lem:htcc:reduct}
 For any interpretation $\langle h,t \rangle$ and formula $\varphi$,
 we have $\langle h,t \rangle\modelsf\varphi$ iff $h\modelscl\varphi^t$
\end{lemma}

\begin{proof}{Lemma~\ref{lem:htcc:reduct}}
    We proof Lemma~\ref{lem:htcc:reduct} via induction over the structure of the formula.
    Lemma~\ref{lem:htcc:confree} constitutes the induction base of this proof.
   \paragraph{Induction step} 
   For $\varphi=(\varphi_1 \otimes \varphi_2) \text{ and } \otimes \in \{\land,\lor,\rightarrow\}$, 
   we refer to the respective induction steps in the proof of Lemma~\ref{lem:htcc:confree}
   since they hold regardless of whether $\varphi_1$ and $\varphi_2$ are condition-free.
    \paragraph{Induction step} For conditional constraint atom $c$,
          we get
      \begin{align}
                & \langle h,t\rangle\modelsf c \label{eq:prop:1} \\
\text{iff } & h\in\den{\evalf{h}{t}(c[\cterm{s_1}{s'_2}{\varphi_1},\cterm{s_2}{s'_2}{\varphi_2},\dots])}\label{eq:prop:2}\\
&\text{ and } \langle t,t \rangle \modelsf c\nonumber  \\
\text{iff } & h\in\den{c[\evalf{h}{t}(\cterm{s_1}{s'_1}{\varphi_1}),\evalf{h}{t}(\cterm{s_2}{s'_2}{\varphi_2}),\dots]}\label{eq:propht}\\
&\text{ and }  \langle t,t \rangle \modelsf c\nonumber \\
\text{iff } & h\in\den{c[\evalcl{h}(\cterm{s_1}{s'_1}{\varphi_1^t}),\evalcl{h}(\cterm{s_2}{s'_2}{\varphi_2^t}),\dots]}\label{eq:propfer}\\
&\text{ and } t\modelscl c \nonumber\\
\text{iff } & h\in\den{\evalcl{h}(c[\cterm{s_1}{s'_1}{\varphi_1},\cterm{s_2}{s'_2}{\varphi_2},\dots]^t)}\label{eq:prop:3}\\
&\text{ and } t\modelscl c  \nonumber\\
\text{iff } & h\modelscl c^t\text{ and }t\modelscl c  \label{eq:prop:4}\\
\text{iff } & h\modelscl c^t \label{eq:prop:5}
      \end{align}
      where $\cterm{s_i}{s'_i}{\varphi_i}$ are all conditional terms occuring in $c$. 
    Equivalence between \eqref{eq:prop:1} and \eqref{eq:prop:2} holds by definition of the satisfaction relation.
    Equivalence between \eqref{eq:prop:2} and \eqref{eq:propht} holds by definition of the evaluation function.
    The equivalence between \eqref{eq:propht} and \eqref{eq:propfer} holds, 
	  first, by induction hypothesis since  
	  \begin{align*}
	  &c[\evalf{h}{t}(\cterm{s_1}{s'_1}{\varphi_1}),\evalf{h}{t}(\cterm{s_2}{s'_2}{\varphi_2}),\dots]=\\
	  &\quad c[\evalcl{h}(\cterm{s_1}{s'_1}{\varphi_1^t}),\evalcl{h}(\cterm{s_2}{s'_2}{\varphi_2^t}),\dots]
	  \end{align*}
	  in case that $\langle h,t \rangle\modelsf\varphi_i$ iff $h\modelscl\varphi_i^t$ for $1\leq i \leq n$,
	  and second, since $\langle t,t \rangle \modelsf c$ iff $t\modelscl c$.
    Equivalence between \eqref{eq:propfer} and \eqref{eq:prop:3} holds by definition of the evaluation function.
    Equivalence between \eqref{eq:prop:3} and \eqref{eq:prop:4} holds by definition of the classical satisfaction relation.
    Equivalence between \eqref{eq:prop:4} and \eqref{eq:prop:5} holds by definitions of reduct and classical satisfaction relation
          since if $h\modelscl c^t$ implies $t\modelscl c$ no interpretation satisfies $\bot$.
\end{proof}
\begin{proof}{Theorem~\ref{thm:htcc:eqstab}}
Theorem~\ref{thm:htcc:eqstab} follows directly from Lemma~\ref{lem:htcc:reduct}.
If valuation $t$ is a $\f$-stable model of $\varphi$, 
we know that $\langle t,t \rangle\modelsf\varphi$ and, by Lemma~\ref{lem:htcc:reduct},
$t\modelscl\varphi^t$.
Furthermore, there cannot exist a valuation $h\subset t$ 
such that $h\modelscl \varphi^t$ 
since that would imply $\langle h,t \rangle \modelsf \varphi$ again by Lemma~\ref{lem:htcc:reduct},
which cannot hold since valuation $t$ is a $\f$-stable model of $\varphi$.
Thus, valuation $t$ is a subset minimal model of $\varphi^t$ making it a $\ff$ stable model of $\varphi$.
Similarly, if valuation $t$ is a \ff stable model of $\varphi$, 
we know that $t\modelscl \varphi^t$ and $\langle t,t \rangle\modelsf\varphi$ by Lemma~\ref{lem:htcc:reduct}.
Again, there cannot exist an $h\subset t$ such that $\langle h,t \rangle\modelsf\varphi$, 
because it implies $h\modelscl\varphi^t$ by Lemma~\ref{lem:htcc:reduct},
which cannot hold since $t$ is a \ff stable model of $\varphi$.
In conclusion, 
$\langle t,t \rangle \modelsf \varphi$ 
and there exists no $h\subset t$ such that $\langle h,t \rangle\models\varphi$,
therefore, valuation $t$ is a \f-stable model of $\varphi$.
\end{proof}

\end{document}